\documentclass[11pt]{article}
\usepackage[utf8]{inputenc}
\usepackage{jheppub}
\usepackage{amsmath,amssymb,amsfonts,graphicx,slashed,color,amsthm,mathtools,upgreek, enumerate, tensor, scalerel, subfig}
\usepackage{arydshln}
\usepackage[dvipsnames]{xcolor}
\usepackage{bbold}
\allowdisplaybreaks

\newcommand{\bra}[1]{\langle #1 \vert}
\newcommand{\ket}[1]{\vert #1 \rangle}
\newcommand{\inner}[2]{\langle #1 \vert #2 \rangle}
\newcommand{\avg}[1]{\langle #1 \rangle}

\newcommand{\JT}{{\text{JT}}}
\newcommand{\BR}{{\text{BR}}}
\newcommand{\br}{{\text{br}}}
\newcommand{\bC}{\mathbb{C}}
\newcommand{\bR}{\mathbb{R}}
\newcommand{\cC}{\mathcal{C}}
\newcommand{\cM}{\mathcal{M}}
\newcommand{\cH}{\mathcal{H}}
\newcommand{\cO}{\mathcal{O}}
\newcommand{\cW}{\mathcal{W}}
\newcommand{\tW}{\widetilde{W}}
\renewcommand{\d}{\mathrm{d}}
\renewcommand{\i}{\mathrm{i}}
\newcommand{\D}{\mathrm{D}}
\DeclareMathOperator{\Tr}{Tr}

\theoremstyle{plain}
\newtheorem*{thm-restate}{Theorem \ref{thm:qms_exact}}

\title{Non-Isometric Quantum Error Correction in Gravity}

\author{Arjun Kar}

\affiliation{Department of Physics and Astronomy,
University of British Columbia,\\ 6224 Agricultural Road, Vancouver, BC V6T 1Z1, Canada}

\emailAdd{arjunkar@phas.ubc.ca}

\abstract{We construct and study an ensemble of non-isometric error correcting codes in a toy model of an evaporating black hole in two-dimensional dilaton gravity.
In the preferred bases of Euclidean path integral states in the bulk and Hamiltonian eigenstates in the boundary, the encoding map is proportional to a linear transformation with independent complex Gaussian random entries of zero mean and unit variance.
Using measure concentration, we show that the typical such code is very likely to preserve pairwise inner products in a set $S$ of states that can be subexponentially large in the microcanonical Hilbert space dimension of the black hole. 
The size of this set also serves as an upper limit on the bulk effective field theory Hilbert space dimension.
Similar techniques are used to demonstrate the existence of state-specific reconstructions of $S$-preserving code space unitary operators.
State-specific reconstructions on subspaces exist when they are expected to by entanglement wedge reconstruction.
We comment on relations to complexity theory and the breakdown of bulk effective field theory.
}

\begin{document}
\maketitle

\section{Introduction}\label{sec:intro}

In holographic formulations of quantum gravity \cite{Maldacena:1997re,Witten:1998qj}, semiclassical spacetime emerges from the quantum information theory of the microscopic degrees of freedom \cite{Wheeler1989-WHEIPQ,VanRaamsdonk:2010pw}.
The first hint of this fundamental role played by information theory appears in the study of gravitational entropy formulas \cite{Bekenstein:1973ur,Hawking:1976de,Ryu:2006bv,Hubeny:2007xt,Faulkner:2013ana,Engelhardt:2014gca,Penington:2019npb,Almheiri:2019psf,Almheiri:2019hni} which express von Neumann entropies in quantum gravity in terms of semiclassical quantities like hypersurface areas and the entropy of quantum fields propagating on fixed backgrounds.

The arguments \cite{Lewkowycz:2013nqa,Faulkner:2013ana,Almheiri:2019qdq,Penington:2019kki,Akers:2020pmf} which lead to these formulas often make use of the Euclidean gravitational path integral, a somewhat mysterious object which is semiclassical at first sight but has been known nearly since its invention \cite{DeWitt:1967ub,Hawking:1980gf} to contain some amount of information about the microscopic degrees of freedom in quantum gravity \cite{Harlow:2021lpu}.\footnote{Lorentzian approaches to gravitational entropy have historically relied upon a string theoretic or holographic formulation of quantum gravity where the underlying degrees of freedom are explicitly known \cite{Cardy:1986ie,Strominger:1996sh,Callan:1996dv,Strominger:1997eq,Maldacena:1997de,Hartman:2014oaa}.  However, recent progress in algebraic quantum field theory has supplied a new perspective on this issue, one that may be as general and independent of stringy details as the Euclidean path integral \cite{Chandrasekaran:2022eqq}.}
Understanding precisely how much microscopic information is accessible via these Euclidean techniques is a very active area of inquiry \cite{Saad:2019lba,Cotler:2021cqa,Iliesiu:2022kny} in light of recent advances concerning the application of some of the more powerful of the aforementioned entropy formulas to the black hole information paradox \cite{Penington:2019npb,Almheiri:2019psf,Almheiri:2019hni,Almheiri:2019qdq,Penington:2019kki,Balasubramanian:2020hfs}.\footnote{The generality and apparent utility of these formulas has even led to efforts to reproduce \cite{Balasubramanian:2020coy,Balasubramanian:2020xqf,Balasubramanian:2021wgd} and analyze \cite{Hartman:2020khs,Hartman:2020swn,Susskind:2021esx,Shaghoulian:2021cef,Shaghoulian:2022fop,Levine:2022wos,Chandrasekaran:2022cip} them in gravitational theories which are not known to be holographic.}

On their own, such entropy formulas and Euclidean gravity calculations have taught us much about the structure of quantum information in quantum gravity (see \cite{Hayden:2011ag,Wall:2012uf,Swingle:2012wq,Headrick:2014cta,Freedman:2016zud} for just a few examples).
But perhaps their deepest implication lies in their relationship with quantum error correcting codes \cite{Almheiri:2014lwa,Jafferis:2015del,Dong:2016eik,Harlow:2016vwg,Akers:2021fut}.
Quantum error correcting codes were originally created to allow for robust manipulation and transmission of quantum information \cite{Shor:1995oct}, and their very existence is somewhat surprising due to the no-cloning principle of quantum mechanics.

As a consequence of the entropy formulas, a quantum error correcting structure was discovered in holography \cite{Almheiri:2014lwa} which consists of a linear map $V$ that encodes a bulk semiclassical ``code'' Hilbert space $\cH_b$ into the microscopic ``physical'' boundary Hilbert space $\cH_B$.
The basic idea is that semiclassical bulk states and operators on $\cH_b$ are encoded redundantly by $V$ in the microscopic Hilbert space $\cH_B$, and losing access to some portions of $\cH_B$ does not necessarily obstruct our ability to reconstruct bulk physics in some portions of $\cH_b$.
The portions in question are determined by the hypersurfaces appearing in the gravitational entropy formulas we recalled above \cite{Headrick:2014cta,Dong:2016eik}, and this phenomenon is sometimes called ``entanglement wedge reconstruction''.

Since its discovery in this fashion, the error correction structure in holography has come to essentially supersede the holographic gravitational entropy formulas, as these formulas and their generalizations are now understood as direct consequences of this structure \cite{Harlow:2016vwg,Akers:2021fut}.
As such, the study of gravitational entropy is superseded by the study of the three fundamental objects in the error correction structure: the code space $\cH_b$, the physical space $\cH_B$, and the encoding map $V$.
As the behavior of gravitational entropy underpins our picture of the emergent bulk spacetime, understanding the limits of semiclassical bulk physics is intimately related to how far this error correction structure may be extended \cite{Hayden:2017xed,Hayden:2018khn,Akers:2021fut,Akers:2022qdl}.

The choice of $\cH_b$ has involved some degree of arbitrariness, and has generally been taken to be a finite subspace of the bulk effective field theory Hilbert space, just large enough to contain the perturbative semiclassical dynamics of interest \cite{Harlow:2018fse}.
On the other hand, $\cH_B$ is almost always taken to be the complete microscopic Hilbert space of the holographic dual boundary theory, perhaps restricted to some fixed charge sector or microcanonical energy window.
The interpretation of $V$ in holography is as the ``bulk-to-boundary map'', and in Euclidean gravity it roughly corresponds to a rule for transforming bulk path integrals into boundary path integrals.

Using the Euclidean gravitational path integral \cite{Almheiri:2014lwa,Jafferis:2015del,Dong:2016eik,Penington:2019kki} or tensor network models of holography \cite{Pastawski:2015qua,Hayden:2016cfa}, it is possible to explicitly define the encoding map $V$ and reconstruct bulk operators acting on the code space $\cH_b$ by manipulating the physical degrees of freedom in $\cH_B$.\footnote{Other bulk reconstruction techniques \cite{Witten:1998qj,Gubser:1998bc,Balasubramanian:1998sn,Banks:1998dd,Susskind:1998dq,Hamilton:2005ju,Hamilton:2006az} make use of the causal structure of spacetime and provide only indirect access to $V$, often without an explicit choice of code space $\cH_b$.  This subtlety led to several puzzles concerning the structure of effective field theory within holography, and unraveling these issues led directly to the quantum error correction ideas we have been reviewing \cite{Almheiri:2014lwa,Harlow:2018fse}.  Modular theory has also been employed as a reconstruction technique \cite{Faulkner:2017vdd} which may be versatile enough to handle choices of $\cH_b$ which include bulk subregions that are separated from $\cH_B$ by a horizon \cite{Chen:2019iro,Jafferis:2020ora,Gao:2021tzr,Jafferis:2022toa}.  }
In these explicit situations, we will loosely refer to $V$ as the holographic dictionary, despite its apparent difference from the original notion of the holographic dictionary where (noting the lack of a non-perturbatively defined bulk theory) $V$ was instead thought of as an isomorphism from a bulk string theory Hilbert space to a boundary gauge theory Hilbert space \cite{Witten:1998qj,Gubser:1998bc,Balasubramanian:1998sn}.

The standard situation in quantum error correction is for the encoding map $V$ to be an isometry.
This means $V$ satisfies $V^\dagger V = I_b$, where $I_b$ is the identity on $\cH_b$.
In some toy models and many holographic scenarios, the dictionary $V$ really is an isometry and leads to an approximate quantum error correcting code between the bulk and boundary where all states in $\cH_b$ look roughly the same from a semiclassical perspective.\footnote{Technically, this means that bulk subregions are encoded in boundary subregions for every state in $\cH_b$ in the same manner.  This is sometimes called ``complementary recovery'' in error correction or ``subregion duality'' in holography.}

However, extensions of the standard error correction structure are necessary and indeed sufficient to understand more precise notions of information in bulk subregions \cite{Hayden:2017xed,Hayden:2018khn}, choices of $\cH_b$ containing states with highly dissimilar bulk entanglement structure \cite{Akers:2020pmf}, and even certain extreme situations motivated by the interiors of evaporating or old black holes where $V$ can be arbitrarily far from an isometry \cite{Kim:2020cds,Akers:2021fut,Balasubramanian:2022fiy,Akers:2022qdl}.\footnote{In the non-isometric situation, the map $V$ may annihilate some states in $\cH_b$.  These null states have been related to a sort of large diffeomorphism invariance \cite{Jafferis:2017tiu,Marolf:2020xie,Balasubramanian:2020jhl}.}
Of these extensions to standard error correction, the non-isometric extension is the least understood and the most relevant for situations with strong gravitational effects.
In the most detailed study to date of this non-isometric error correction \cite{Akers:2022qdl}, the model under consideration was a tensor network model similar in spirit to \cite{Hayden:2016cfa}, with no obvious connection to gravity.

As was already explained in \cite{Akers:2022qdl}, the tensor network model has several drawbacks as a model of gravity.
For example, Lorentz and diffeomorphism invariance are hard to understand in such models, and the quantity which plays the role of the microcanonical black hole Hilbert space dimension has no obvious connection to a geometric area as expected by the Bekenstein-Hawking formula.
On the other hand, the Euclidean path integral naturally allows for manifest diffeomorphism invariance, and the entropy of the black hole is clearly related to the horizon area.
Indeed, computations in Euclidean gravity were the original justification for the Bekenstein-Hawking formula itself.
In view of these facts, it is necessary to understand the extent to which the results obtained by tensor network analysis in \cite{Akers:2022qdl} may be carried over to a real gravitational theory defined using the Euclidean path integral.

The purpose of this paper is to construct and study the dictionary $V$ as a highly non-isometric quantum error correcting code in a toy model of an evaporating black hole in two-dimensional Euclidean dilaton gravity.
We combine the Euclidean gravity techniques which allow for a resolution of the information paradox \cite{Penington:2019kki} with the notion of non-isometric error correction introduced in \cite{Akers:2022qdl}.
We find gravitational analogues of many of the results of \cite{Akers:2022qdl}, with interesting differences in details.
In particular, our analogue \eqref{eq:overlap-bound} of the main theorem of \cite{Akers:2022qdl} dealing with changes in semiclassical state overlaps under the dictionary is strengthened by differences in measure concentration between the Haar ensemble and the complex Gaussian ensemble, which enters our gravitational code.
This strengthening carries over to the bulk reconstruction analyses in e.g.~\eqref{eq:global-state-spec-recon}.
Moreover, the derivations themselves are in fact a bit simpler than in \cite{Akers:2022qdl}, which is surprising as gravitational theories are generally more complicated than Haar random unitary analyses.

Four sections follow.
In Section~\ref{sec:dictionary}, we define the dictionary $V$ in two-dimensional dilaton gravity and briefly review how the model relates to evaporating black holes.
In Section~\ref{sec:overlap}, we prove the gravitational analogue of the fundamental theorem of \cite{Akers:2022qdl}, which allows an estimation of how many states in the code Hilbert space may have preserved overlaps under the dictionary $V$.
In Section~\ref{sec:reconstruction}, we describe the reconstruction of code operators in a necessarily state-specific manner and verify consistency of our results with entanglement wedge reconstruction.
We conclude in Section~\ref{sec:disc} with a discussion of relations to complexity, fundamental averaging, extensions to gravitational bulk operators, and the breakdown of bulk effective field theory.

\section{Holographic dictionary}\label{sec:dictionary}

We will define the holographic dictionary $V$ using the Euclidean gravitational path integral.
Our discussion will apply quite generally to asymptotically anti-de Sitter (AdS) two-dimensional dilaton gravity theories, but for concreteness we will also spell out the details for a particular theory: Jackiw-Teitelboim (JT) gravity \cite{Teitelboim:1983ux,Jackiw:1984je}, which is a theory of the metric $g_{ab}$ and a single real scalar dilaton field $\phi$ with Euclidean action
\begin{equation}
\begin{split}
    I_\JT(\cM,g_{ab},\phi) = & -\frac{S_0}{2\pi} \left[ \frac{1}{2} \int_\cM \sqrt{g} R + \int_{\partial \cM} \sqrt{h} K \right] \\
    & - \left[ \frac{1}{2} \int_\cM \sqrt{g} \phi (R+2) + \int_{\partial \cM} \sqrt{h}\phi (K-1) \right]\ ,
\end{split}
\label{eq:jt-action}
\end{equation}
and asymptotic boundary conditions
\begin{equation}
    \d s^2|_{\partial \cM} = \frac{\d \tau^2}{\epsilon^2}\ , \quad \phi|_{\partial \cM} = \frac{1}{\epsilon}\ , \quad \epsilon \to 0\ .
\label{eq:jt-boundary-conds}
\end{equation}

The basic idea, following \cite{Penington:2019kki}, is to consider a set of independent end-of-the-world brane states in the bulk \cite{Kourkoulou:2017zaj} which corresponds to a boundary ensemble of pure states with independent complex Gaussian random coefficients in the energy eigenbasis.
In the microcanonical ensemble, the smooth energy and brane tension dependence of these states is nearly eliminated, and we are left with an ensemble of microcanonical holographic dictionaries consisting of (writing $|A| = \dim \cH_A$) rectangular matrices of dimension $|B| \times |b|$ with independent Gaussian random entries of zero mean and unit variance.
This ensemble of dictionaries is therefore capturing a kind of topological sector of holographic dictionaries which is largely independent of the underlying dynamics.
We now turn to deriving this ensemble from Euclidean gravity.

\subsection{Canonical ensemble}

In the microscopic boundary theory with Hamiltonian $H$, we may begin with the path integral that computes the thermal partition function
\begin{equation}
    Z(\beta) = \Tr e^{-\beta H}\ .
\end{equation}
We may represent this trace with a path integral on a ``thermal circle'' $S^1(\beta)$ of length $\beta$ with periodic coordinate $\tau \sim \tau + \beta$.
\begin{figure}
    \centering
    \includegraphics[scale=.4]{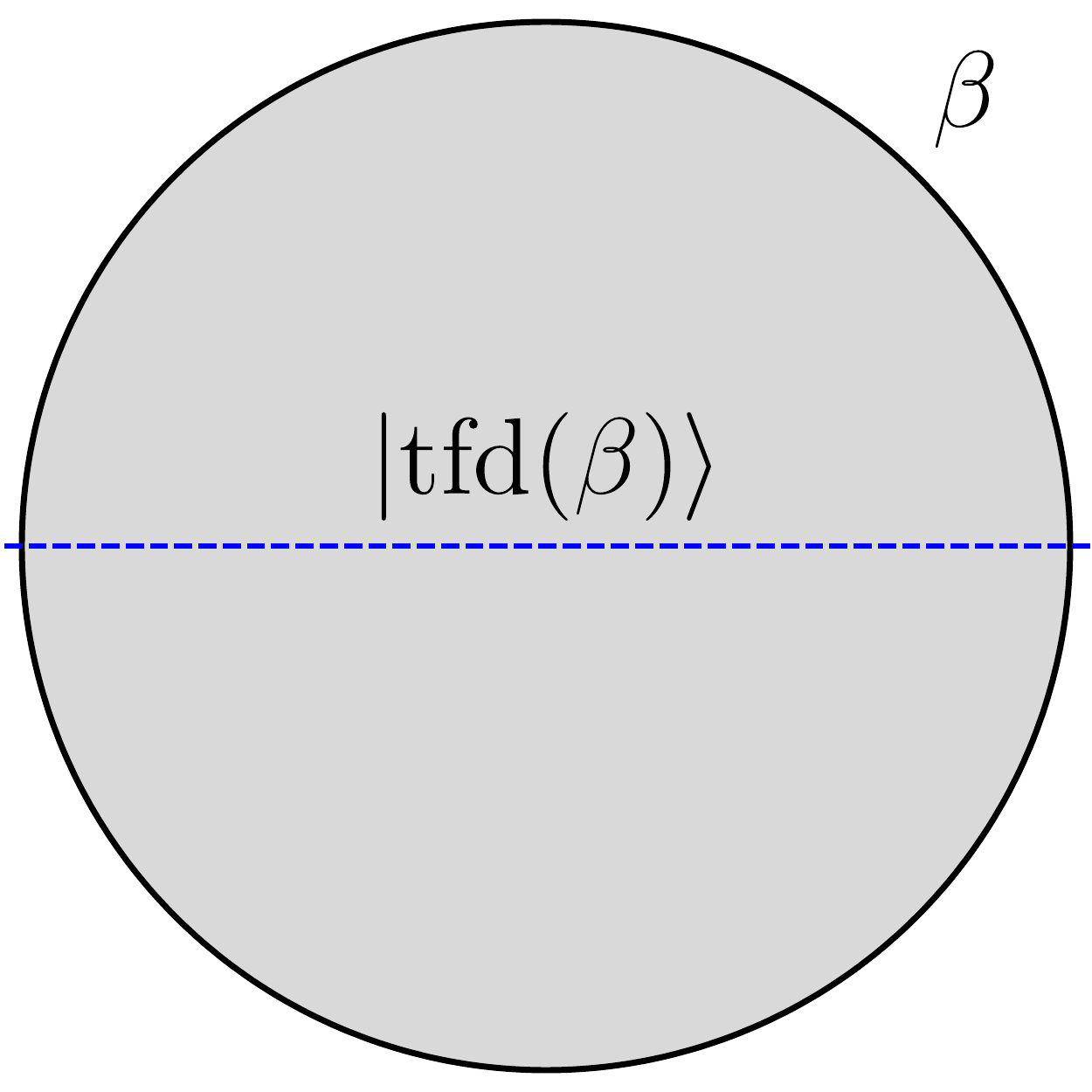}
    \hspace{1.5cm}
    \includegraphics[scale=.4]{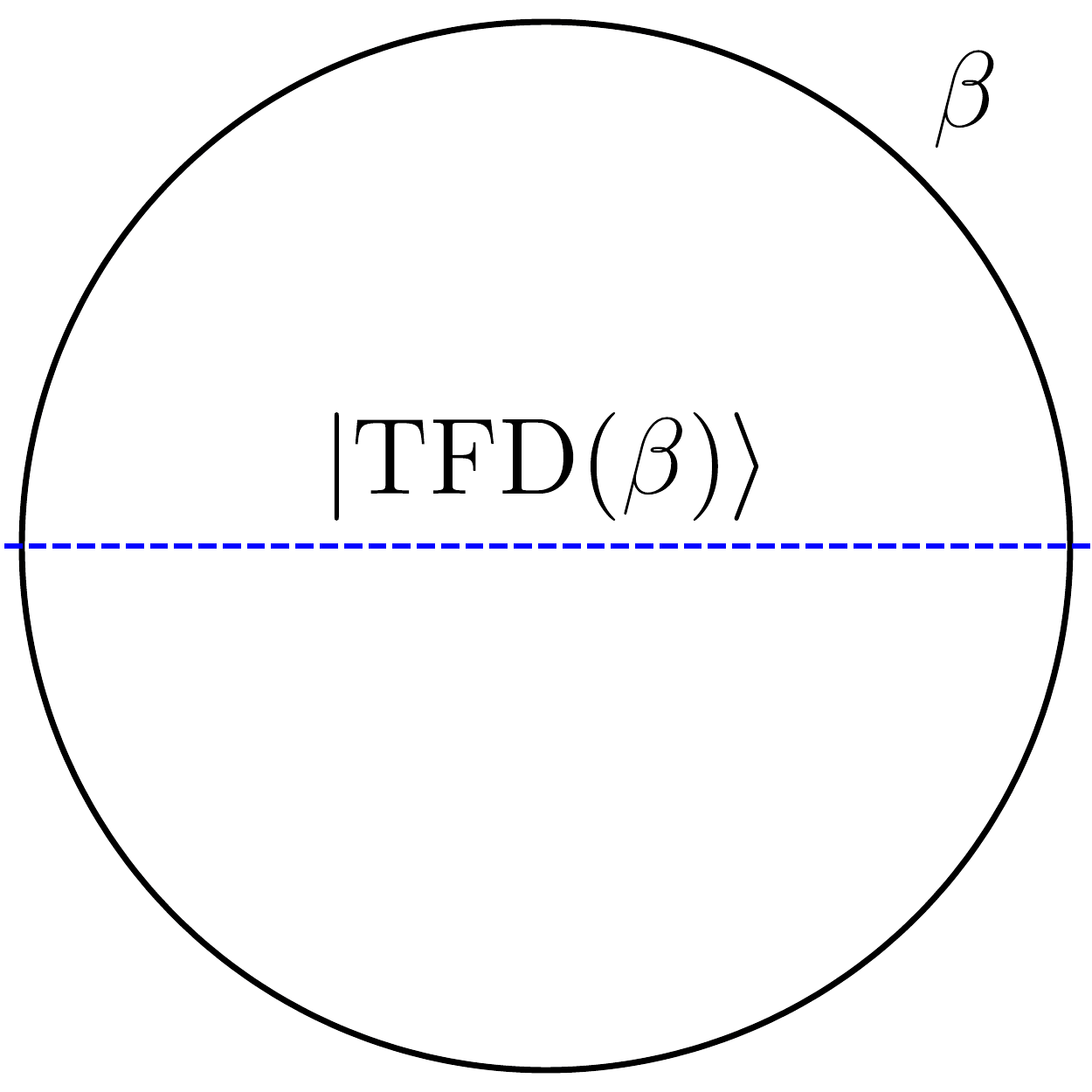}
    \caption{
    \small{
    \textbf{(Left)} The disk contribution to the bulk Euclidean path integral computing the thermal partition function.  The boundary has renormalized length $\beta$. Cutting the path integral along the dashed blue line gives a state in $\cH_\text{grav}$, the Hilbert space of dilaton gravity on an interval with two asymptotic boundaries.
    This state is the eternal two-sided black hole in AdS$_2$.
    \textbf{(Right)} The boundary Euclidean path integral computing the thermal partition function.  Cutting the path integral along the dashed blue line gives the thermofield double state in two copies $\cH_B \otimes \cH_B$ of the microscopic Hilbert space.
    In the asymptotic $e^{-S_0}$ expansion of JT gravity \eqref{eq:jt-action}, this path integral is computed by summing over all possible Euclidean geometries and topologies which fill in the circle.  This includes the disk, the disk with a handle, and so on.
    }
    }
    \label{fig:tfd}
\end{figure}
Holographically, this path integral is computed by summing over all Euclidean geometries and topologies which fill in this thermal circle with the boundary conditions \eqref{eq:jt-boundary-conds}, weighted by the exponential of minus the bulk gravitational action.
For small enough inverse temperature $\beta$, the dual bulk path integral is dominated\footnote{Unlike in higher dimensions, there is no Hawking-Page transition \cite{Hawking:1982dh} in AdS$_2$.  However, there is an ultra-low temperature regime where pathological thermodynamic behavior can be observed unless higher topologies (and eventually the full non-perturbative completion) are included \cite{Engelhardt:2020qpv}.} by the $\cM = D^2$ Euclidean black hole geometry where the thermal circle is filled in with a hyperbolic disk.

Cutting this path integral in half on the boundary, we obtain a quantum state in two copies of the boundary Hilbert space.
Explicitly, it is the (unnormalized) thermofield double state
\begin{equation}
    \ket{\text{TFD}(\beta)} \equiv \sum_{n=0}^\infty e^{-\beta E_n/2} \ket{E_n} \otimes \ket{E_n}\ ,
\label{eq:TFD}
\end{equation}
where $E_n$ are the eigenvalues of the boundary Hamiltonian $H$ on the orthonormal eigenstates $\ket{E_n} \in \cH_B$.
In the bulk, cutting the disk topology in half with the saddle point metric yields the eternal two-sided AdS black hole geometry \cite{Maldacena:2001kr}, and the bulk Cauchy slice has the interval topology.
The semiclassical approximation to \eqref{eq:TFD} is therefore a state $\ket{\text{tfd}(\beta)} \in \cH_\text{grav}$ created by summing over geometries with fixed half-disk topology (Figure~\ref{fig:tfd}).
The Hilbert space $\cH_\text{grav}$ is the canonically quantized Hilbert space of the gravity theory on the interval Cauchy slice with two asymptotic boundaries.\footnote{This Hilbert space has been constructed by canonical quantization in two-dimensional dilaton gravity theories \cite{Louis-Martinez:1993bge,Cavaglia:1998xj,Harlow:2018tqv}, but more generally it is difficult to define explicitly.  An appropriate formal prescription for defining $\cH_\text{grav}$ more generally might be found in algebraic field theory, where a Hilbert space description of perturbative quantum gravity has been developed recently using the GNS construction and crossed product from the theory of von Neumann algebras \cite{Leutheusser:2021frk,Leutheusser:2021qhd,Witten:2021jzq,Witten:2021unn}.}
Schematically, this state is given by
\begin{equation}
    \ket{\text{tfd}(\beta)} = \int_0^\infty \d E\; \rho(E) e^{-\beta E/2} \ket{E}_\text{grav}\ ,
\label{eq:semi-TFD}
\end{equation}
where $\ket{E}_\text{grav}$ is a basis for $\cH_\text{grav}$ with fixed asymptotic energy $E$ measured from either end of the interval Cauchy slice and overlap
\begin{equation}
    \inner{E}{E'}_\text{grav} = \frac{\delta(E-E')}{\rho(E)}\ ,
\end{equation}
and $\rho(E)$ is the density of states implied by approximating the thermal partition function $Z(\beta)$ with the bulk disk topology.

Generically, the exact disk path integral will yield a smooth function of energy for $\rho(E)$ rather than a quantum mechanical density of states, which would instead be a sum of Dirac delta functions centered on the eigenvalues of the Hamiltonian.\footnote{See \cite{Blommaert:2021fob,Blommaert:2022ucs} for finely tuned dilaton gravity theories where the exact disk path integral, plus a few corrections from additional nonlocal degrees of freedom, does yield a discrete density of states.  We will avoid introducing such additional degrees of freedom here.}
In JT gravity \eqref{eq:jt-action}, the path integral on the disk topology can be evaluated exactly \cite{Maldacena:2016upp,Stanford:2017thb}, and the corresponding density is
\begin{equation}
    \rho_\JT(E) = \frac{1}{2\pi^2} \sinh (2\pi\sqrt{2E})\ .
\end{equation}

With these states in hand, we are ready to write the simplest relation involving the canonical holographic dictionary:
\begin{equation}
    V \ket{\text{tfd}(\beta)} = \ket{\text{TFD}(\beta)}\ .
\end{equation}
We can roughly think of the dictionary $V$ as ``hollowing out'' the bulk path integral and leaving behind the boundary conditions, which are enough to define the boundary path integral.

\subsection{End-of-the-world branes}

The dictionary we are interested in is supposed to yield a state in a single copy of $\cH_B$.
But the microscopic thermofield double \eqref{eq:TFD} lives in two copies of $\cH_B$.
To produce a state in the correct Hilbert space, we need to project away one of the Hilbert space factors from $\ket{\text{TFD}(\beta)}$.

However, we must take care not to destroy too much of the bulk semiclassical structure by this projection.
If we are not careful, we may disrupt the entire spacetime beyond the event horizon of the remaining side, effectively eliminating the interior.
This is not a problem in principle for the definition of the dictionary $V$, but in Section~\ref{sec:reconstruction} we will discuss the reconstruction of operators in the interior, and the lack of any smooth interior region at all would pose a problem for the definition of the operators we will try to reconstruct.

Kourkoulou and Maldacena described a fairly straightforward method to implement a projection while preserving a sizable smooth region behind the horizon \cite{Maldacena:2001kr,Kourkoulou:2017zaj}.
We augment the Euclidean bulk gravity action with end-of-the-world branes via
\begin{equation}
    I_\text{grav} \to I_\text{grav} - \int_\text{brane} \d s\ (\phi K - \mu ) \ ,
\end{equation}
where the integration measure is the proper length element on the brane\footnote{
This length is divergent in AdS and requires regularization.  A standard method is to subtract a simple function of the dilaton which diverges at the same rate at the boundary $\partial \cM$.} and $\mu$ is the brane tension.
For the moment, we have only a single type of brane, but we will introduce multiple ``flavors'' of branes shortly which do not interact in the bulk, and this extension just corresponds to adding multiple brane terms in the action which are to be integrated only along branes of the appropriate flavor.
On the brane, we enforce ``dual'' boundary conditions
\begin{equation}
    K = 0\ , \quad n^a \partial_a \phi = \mu\ ,
\end{equation}
where $K$ is the extrinsic curvature and $n^a$ are the coefficients of the vector normal to the brane.

\begin{figure}
    \centering
    \includegraphics[scale=.4]{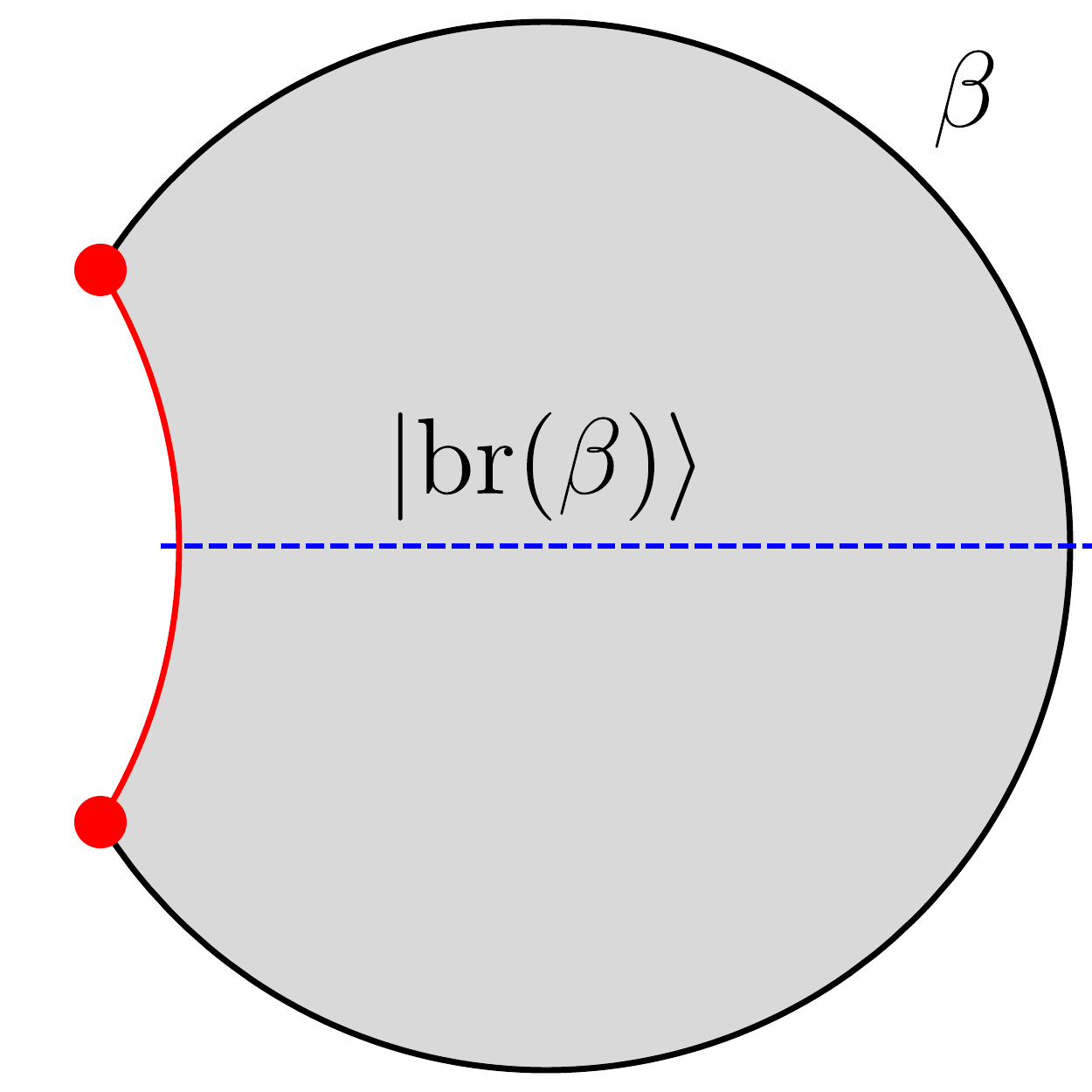}
    \hspace{1.5cm}
    \includegraphics[scale=.4]{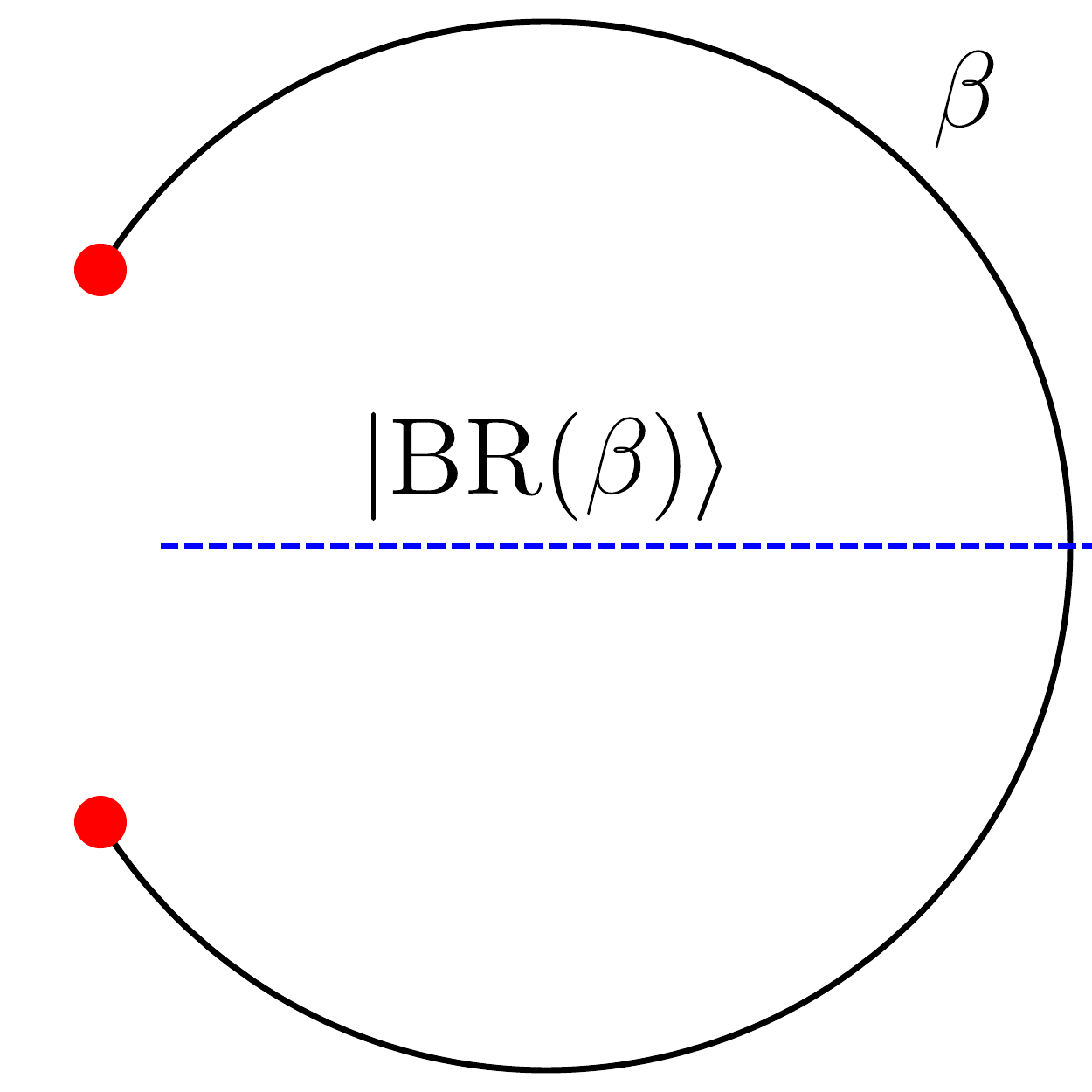}
    \caption{
    \small{
    \textbf{(Left)} The disk contribution to the bulk Euclidean path integral computing the one-brane (red curve) partition function.  The boundary has renormalized length $\beta$. Cutting the path integral along the dashed blue line gives a state in $\cH_\text{br}$, the Hilbert space of dilaton gravity on an interval with one asymptotic boundary and one brane boundary.
    \textbf{(Right)} The boundary Euclidean path integral computing the one-brane partition function.  Cutting the path integral along the dashed blue line gives the projected thermofield double state $\inner{S_\mu}{\text{TFD}(\beta)}$, which we call $\ket{\BR(\beta)}$ in \eqref{eq:BR}, in the microscopic Hilbert space $\cH_B$.
    The red dots are brane boundary conditions.
    }
    }
    \label{fig:br}
\end{figure}

In the disk topology, the brane connects with the thermal circle at two points and passes through the bulk Cauchy slice orthogonally (Figure~\ref{fig:br}).
As such, the bulk Cauchy slice topology is still an interval, but with only one asymptotic boundary.
The second asymptotic boundary is cut off by the brane, which is at a finite distance from the bifurcation surface of the eternal black hole.
The boundary Cauchy slice, on the other hand, is now a single point and the semicircular contour creating $\ket{\text{TFD}(\beta)}$ becomes instead a segment with a brane boundary condition at one end.

This brane boundary condition is a microscopic projection operator which we may think of as a state $\ket{S_\mu} \in \cH_B$, and the boundary Euclidean path integral now yields a state
\begin{equation}
    \ket{\BR(\beta)} \equiv \inner{S_\mu}{\text{TFD}(\beta)} = \sum_{n=0}^\infty f_\mu(E_n) e^{-\beta E_n/2} C_n \ket{E_n}\ ,
\label{eq:BR}
\end{equation}
where we have defined $f_\mu (E_n) C_n \equiv \inner{S_\mu}{E_n}$.
There is a corresponding semiclassical state in $\cH_\br$, the canonical Hilbert space of dilaton gravity on an interval with one asymptotic and one brane boundary, which is obtained by restricting again to the disk topology, and this yields
\begin{equation}
    \ket{\br(\beta)} = \int_0^\infty \d E\; f_\mu (E) \rho(E) e^{-\beta E/2} \ket{E}_\br\ ,
\label{eq:semi-BR}
\end{equation}
where the state $\ket{E}_\br$ is again labeled by the asymptotic energy and has the same normalization as $\ket{E}_\text{grav}$.\footnote{As with \eqref{eq:semi-TFD}, the single-sided Hilbert space $\cH_\br$ arising from canonical quantization with one asymptotic boundary and one brane boundary is known to exist in JT gravity with the fixed asymptotic energy states forming a complete basis \cite{Gao:2021uro}.  To our knowledge, the analogous result for a general dilaton gravity theory has not been demonstrated, but we consider it plausible in view of \cite{Louis-Martinez:1993bge,Cavaglia:1998xj}. }

The function $f_\mu (E_n)$ captures the tension and energy dependence of the overlap between the brane boundary condition state and the Hamiltonian eigenstate.
In JT gravity \eqref{eq:jt-action}, it has the compact expression \cite{Gao:2021uro}
\begin{equation}
    f^\JT_\mu (E) = \; \Gamma \left( \mu + \frac{1}{2} + \i \sqrt{2E} \right)\ ,
\end{equation}
where we have ignored overall constant factors for simplicity.
The role of the coefficients $C_n$ in \eqref{eq:BR} is more subtle and much more crucial for our construction.
It turns out that the $C_n$ are independent complex Gaussian random variables with zero mean and unit variance 
\begin{equation}
    C_n \sim \bC\text{Normal}(0,1)\ ,
\end{equation}
and extending the bulk theory by end-of-the-world branes as we have done actually introduces an ensemble of holographic dual theories \cite{Penington:2019kki}.\footnote{In fact, in JT gravity \cite{Saad:2019lba} and more general dilaton gravity theories \cite{Stanford:2019vob,Witten:2020wvy,Maxfield:2020ale}, the boundary Hamiltonian $H$ itself is a random matrix.  See \cite{Eynard:2015aea} for a mathematical introduction and Section 2 of \cite{Johnson:2020lns} or Appendix C of \cite{Kar:2022sdc} for a more gravity-focused treatment.  The random matrix structure of dilaton gravity will not be important for us, so we will not dwell on it any further.}

The presence of this ensemble is more clear when we consider the problem of computing the microscopic overlap product $\inner{\BR(\beta_1)}{\BR(\beta_1)} \dots \inner{\BR(\beta_n)}{\BR(\beta_n)}$.
This boundary expression is computed holographically by a gravitational path integral with $n$ asymptotic boundary segments with renormalized lengths $\beta_1, \dots , \beta_n$, each with two endpoints from which oriented branes will propagate into the bulk.
The orientation is such that a brane must exit from an endpoint associated with a ket, and enter an endpoint associated with a bra.
The ensemble average over the complex Gaussian coefficients implements precisely the necessary Kronecker delta functions to capture all possible allowed patterns of brane connections through the bulk.

The fact that there is an independent coefficient for each energy eigenstate corresponds to the bulk property that all asymptotic regions in a given connected geometry must have the same asymptotic energy.
Roughly speaking, this property holds because if two asymptotic boundaries are connected through the bulk, there exists a Cauchy slice connecting them with the same asymptotic energy at both ends; this is nothing but the $\ket{E}_\text{grav}$ basis from earlier.
So the ensemble average also acts to match the energy sums in all asymptotic segments that bound a given connected bulk region.
As we discussed, the $\ket{E}_\text{grav}$ basis exists quite generally in dilaton gravity theories, so the complex Gaussian ensemble of brane states appears in any such theory also.\footnote{In this way, our construction is roughly as general as the analysis in \cite{Stanford:2020wkf} of certain probe matter-supported wormhole solutions in dilaton gravity.
The discussion there relied on both the $\ket{E}_\text{grav}$ basis as well as the geodesic length basis $\ket{\ell}_\text{grav}$ in which probe operators were defined.}

Using \eqref{eq:BR} and \eqref{eq:semi-BR}, we are ready to write another entry in the canonical holographic dictionary, this time relating the semiclassical and microscopic brane states via
\begin{equation}
    V \ket{\br(\beta)} = \ket{\BR(\beta)}\ .
\label{eq:dictionary-abstract}
\end{equation}

\subsection{Microcanonical ensemble}

We have thus far described the canonical ensemble where the renormalized proper lengths of asymptotic boundaries are fixed.
For our analysis in Section~\ref{sec:overlap} and beyond, we are more interested in the microcanonical ensemble where the dimension of $\cH_B$ is finite.
There is a simple way to translate formulas from the canonical ensemble to the microcanonical ensemble.
In the microcanonical ensemble, we consider states which have energies in a small band $(E, E+\Delta E)$.
Thus, we make the replacements
\begin{align}
    \int_0^\infty \d E\; \rho (E) & \to \rho (E) \Delta E\ , \label{eq:microcanonical-integral} \\
    \sum_{n=0}^\infty f_\mu (E_n) e^{-\beta E_n/2} \ket{E_n} & \to f_\mu(E) e^{-\beta E/2} \sum_{E_n=E}^{E+\Delta E} \ket{E_n} \ , \label{eq:microcanonical-sum}
\end{align}
as $E$ is now held approximately fixed at the asymptotic boundary, and $e^{-\beta E}$, $f_\mu (E)$, and $\rho (E)$ are varying on scales much larger than $\Delta E$ and therefore are effectively constant.

This leads to a microcanonical entropy $\textbf{S}(E)$, a microcanonical one-brane partition function $\textbf{Z}_1(E)$, and a (normalized) microcanonical state $\ket{\textbf{\BR}(E)}$ given by
\begin{align}
    \exp {\textbf{S}(E)} & \equiv \rho (E) \Delta E\ ,\\ 
    \textbf{Z}_1(E) & \equiv \rho (E) e^{-\beta E} |f_\mu(E)|^2 \Delta E\ ,\label{eq:microcanonical-normalization}\\
    \ket{\textbf{\BR}(E)} & \equiv \frac{1}{\sqrt{\textbf{Z}_1(E)}} e^{-\beta E/2} f_\mu(E) \sum_{E_n=E}^{E+\Delta E} C_n \ket{E_n}\ .\label{eq:microcanonical-brane-state}
\end{align}
The state \eqref{eq:microcanonical-brane-state} is the microcanonical analogue of the canonical ensemble state \eqref{eq:BR}, and $\textbf{Z}_1(E)$ is the microcanonical version of the disk path integral on the left in Figure~\ref{fig:br}.
Simplifying \eqref{eq:microcanonical-brane-state} and relabeling the energies in the microcanonical window, we have
\begin{equation}
     \ket{\textbf{\BR}(E)} = \exp \left(-\frac{\textbf{S}(E)+\i \eta_\mu(E)}{2} \right) \sum_{n=1}^{\exp \textbf{S}(E)} C_n \ket{E_n} \ ,
\end{equation}
where we have also defined the phase
\begin{equation}
    \exp (-\i \eta_\mu(E)/2) \equiv \frac{f_\mu(E)}{|f_\mu(E)|}\ .
\end{equation}

Crucially, the only remaining dependence on the microscopic dynamics is via the eigenstates $\ket{E_n}$.
But in this work, we will be most interested in computing norms and overlaps in the microscopic theory using the energy basis, and we know the orthonormal inner product formula $\inner{E_n}{E_m} = \delta_{mn}$ holds.
So, if we are careful to only compute inner products in the energy basis then the microscopic dynamics plays no further role in the dictionary and we may replace the eigenstates $\ket{E_n}$ with a fixed orthonormal basis $\ket{n}$ of an abstract Hilbert space $\cH_B$ with dimension $e^{\textbf{S}(E)}$.\footnote{This also works for theories with ensemble duals.  For instance, the orthonormality of energy eigenstates holds in every member of the Hamiltonian ensemble dual to JT gravity, so if we always compute overlaps in the energy basis we can treat abstractly the particular states $\ket{E_n}$ in the matrix model Hilbert space.}
The phase $\eta_\mu(E)$ will also not play any further role since we will usually study the combination $V^\dagger V$ where this overall phase will cancel, so we drop it for convenience.

We are left with an ensemble of microcanonical holographic dictionaries determined by a set of complex Gaussian random variables $C_n$ with zero mean and unit variance
\begin{equation}
    V\ket{\textbf{\br}(E)} = \exp({-\textbf{S}(E)/2}) \sum_{n=1}^{\exp \textbf{S}(E)} C_n \ket{n}\ ,
\end{equation}
where $\ket{\textbf{\br}(E)}$ is the normalized microcanonical analogue of \eqref{eq:semi-BR} and $\ket{n}$ is a preferred orthonormal basis basis of an abstract finite Hilbert space
\begin{equation}
    \cH_B \equiv \bC^{\exp \textbf{S}(E)}\ ,
\end{equation}
descending from its origin as a subspace of energy eigenstates.

We may further add more ``flavors'' of branes to the theory which do not interact with each other in the bulk.
This creates superselection sectors which we label by $\alpha = 1,\dots , |b|$, and we have anticipated that the number of flavors we have will correspond precisely to the dimension of a Hilbert space $\cH_b$.
In the path integral, a brane of a particular flavor cannot change flavors, and must begin and end on brane boundary conditions of the appropriate flavor in order for the path integral to be nonzero.
This extends the vector $C_n$ of complex Gaussian random coefficients to a matrix $C_{n \alpha}$ where $\alpha$ labels the brane flavor.

These different flavors are each associated with a distinct orthonormal semiclassical state, and these states form a basis for our code space $\cH_b$.
Formally, we have one copy for each flavor of the gravity Hilbert space $\cH_\br^\alpha$ with a normalized microcanonical state $\ket{\textbf{\br}_\alpha(E)}$, and $\cH_b$ is given by
\begin{equation}
    \cH_b \equiv \oplus_{\alpha = 1}^{|b|} \ket{\textbf{\br}_\alpha(E)}\ ,
\end{equation}
where these states also form a special orthonormal basis of $\cH_b$.
The microcanonical dictionary $V$ acts upon this basis of $\cH_b$ as a complex Gaussian random linear transformation:
\begin{equation}
    V = \exp(-\textbf{S}(E)/2)\ C\ .
\label{eq:dictionary-microcanonical}
\end{equation}
This dictionary, as noted in \cite{Penington:2019kki}, is related to Page's random state model \cite{Page:1993df} with the caveat that it only preserves the norm of vectors on average.
We emphasize that $\cH_b$ is only a subspace of the full canonically quantized bulk gravity Hilbert space with $|b|$ flavors of branes, and we have chosen this subspace in order to simplify the holographic dictionary and the sorts of operators which may act in the bulk.\footnote{A similar choice was made implicitly in \cite{Penington:2019kki}.}
Our choice can be thought of as restricting to a topological sector of the gravity Hilbert space which is controlled only by the brane combinatorics.
Possible extensions of this choice of $\cH_b$ to include more of the gravity Hilbert space along with different types of gravitational operators are discussed in Section~\ref{sec:disc}.

Once we have these brane flavors, we may model an evaporating black hole following \cite{Penington:2019kki} by considering a sequence of states with more and more entanglement between $\cH_b$ and some reservoir system $\cH_R$.
The extended dictionary on $\cH_b \otimes \cH_R$ is taken to be $V \otimes I_R$, using the identity on $R$ to treat the reservoir system equally in both the semiclassical and microscopic pictures.
The brane flavors act like excitations in the interior, and when the entanglement entropy between the interior and the reservoir exceeds the black hole entropy we run into the information paradox \cite{Hawking:1975vcx}.
The paradox is resolved by precisely the non-isometric nature of \eqref{eq:dictionary-microcanonical}, as the dictionary generates small overlaps between states that are orthogonal in the bulk, and this cuts off the reservoir entropy growth at the black hole entropy.

However, unlike the perturbative non-isometry studied in \cite{Penington:2019kki,Balasubramanian:2022fiy} by considering small excitations around such evaporating states, we will consider a much larger portion of $\cH_b \otimes \cH_R$ as a code space.
In doing so, we can try to reconstruct bulk physics in both completely disentangled states and also highly entangled states which cause an information problem.
Precisely how many states admit interesting bulk operator reconstructions is a question we turn to in Section~\ref{sec:overlap}, and we expect that the highly non-isometric nature of \eqref{eq:dictionary-microcanonical} when $|b| \gg |B|$ will prevent us from simply reconstructing all of the code space $\cH_b \otimes \cH_R$.

Before proceeding, we note that it is important for us to write this microcanonical dictionary in the basis defined by the path integral states and energy eigenstates we discussed above.
Only in these bases do we have a simple description of $V$ where each matrix element is an independent random variable.
This is a significant difference compared with \cite{Akers:2022qdl}, where the ensemble of unitary operators defining the code ensemble was itself invariant under a change of basis.
The reason for this difference is the gravitational nature of our code: using geometric path integrals gives a very natural class of states upon which the dictionary simplifies.
The other difference with \cite{Akers:2022qdl} is that our construction has fundamental averaging, while the unitary ensemble was invoked in \cite{Akers:2022qdl} only to study properties of the typical code.
We comment further on this issue in Section~\ref{sec:averaging}. 
Moving forward, we will analyze only the typical code in our fundamental ensemble, using it in the same manner as \cite{Akers:2022qdl} used the unitary ensemble.

\section{Overlap preservation}\label{sec:overlap}

From the discussion in Section~\ref{sec:dictionary} and the final formula \eqref{eq:dictionary-microcanonical}, we saw that the linear transformation
\begin{equation}
    V = |B|^{-1/2}\ C\ ,
\label{eq:dictionary-3}
\end{equation}
where $C: \cH_b \to \cH_B$ is a matrix of complex Gaussian random variables with dimension $|B| \times |b|$, was equivalent to the microcanonical holographic dictionary in dilaton gravity in a certain basis of Euclidean path integral states.
We now want to repeat the analysis of \cite{Akers:2022qdl}, where an ensemble of non-isometric codes was defined using a Haar random unitary transformation instead of a complex Gaussian random transformation like \eqref{eq:dictionary-3}.

The main result driving much of the analysis in \cite{Akers:2022qdl} was a quantitative bound on the size of the set of states in the code space which are expected to have their norms or overlaps approximately preserved by the encoding map.
Using the theory of measure concentration,\footnote{See Appendix B of \cite{Akers:2022qdl} or \cite{Tao:2010mea,Tao:2013bak} for brief introductions.} \cite{Akers:2022qdl} produced a bound on the probability (in the Haar measure on the unitary group) that a typical non-isometric code would modify the norm of a given code space state by more than an exponentially (in $\log |B|$) small amount.
This bound turned out to be doubly exponentially strong (in $\log |B|$), implying by the union bound that a rather large set of states in the code space is expected to have norms and pairwise overlaps preserved by a typical Haar random non-isometric code.

All of this is relevant for understanding what sorts of operators may be reconstructed in a non-isometric code, though we will delay that discussion until Section~\ref{sec:reconstruction}.
For now, we repeat the analysis of \cite{Akers:2022qdl}, using measure concentration for the dictionary ensemble \eqref{eq:dictionary-3} to place a similar doubly exponentially strong bound on the probability that the typical encoding map in \eqref{eq:dictionary-3} modifies the norm of a generic code state by more than an exponentially (in $\log |B|$) small amount.
This will allow us to use the union bound to give a rough estimate of the number of states in the code space which have preserved norms and overlaps under the action of a typical $V$ drawn from \eqref{eq:dictionary-3}.

\subsection{Deviation bound}

We now turn to the question of norm preservation under the encoding map, and in particular we wish to bound the probability (in the complex Gaussian probability measure for $C$) of large deviations in the function
\begin{equation}
    F(C) = \| V \otimes I_R \ket{\psi} \|\ ,
\label{eq:F-defn}
\end{equation}
where $\cH_R$ is an auxiliary space with identity $I_R$, $\ket{\psi} \in \cH_b \otimes \cH_R$, and $\| \cdot \|$ is the Hilbert space norm of $\cH_B \otimes \cH_R$.\footnote{As we reviewed in Section~\ref{sec:dictionary}, in gravitational models $\cH_R$ can be thought of as a radiation bath into which an AdS black hole can evaporate \cite{Penington:2019kki,Balasubramanian:2022fiy}.}
Specifically, for the Gaussian random variables $C$ with respect to the probability measure on $\bC^{|B||b|}$
\begin{equation}
     \D C \equiv \frac{1}{\pi^{|B||b|}}  \exp \left( -\Tr C^\dagger C \right) \bigwedge_{n,\alpha} \frac{ \d C_{n\alpha} \wedge \d\overline{C_{n\alpha}} }{-2\i }\ ,
\label{eq:C-distribution}
\end{equation}
we wish to prove for any $0 < \gamma < \frac{1}{2}$ and any $|B| \geq 4$ that the following bound holds:
\begin{equation}
    \Pr[ |F - 1| \geq |B|^{-\gamma} ] \leq 2 \exp \left( -\frac{|B|^{1-2\gamma}}{2} \right)\ .
\label{eq:dev-bound-F}
\end{equation}

We first obtain a Lipschitz constant $\kappa$ for $F(C)$ under the standard Euclidean distance on $\bC^{|B||b|}$, namely
\begin{equation} 
\d s^2 = \sum_{n,\alpha} \d C_{n\alpha} \d \overline{C_{n\alpha}}\ .
\label{eq:C-metric}
\end{equation}
This means we seek a constant $\kappa$ such that
\begin{equation}
    |F(C_1) - F(C_2)| \leq \kappa \| C_1 - C_2 \|_2\ ,
\label{eq:lipschitz-defn}
\end{equation}
where $\| \cdot \|_p$ is the $p$-norm
\begin{equation}
    \| X \|_p \equiv \left[ \Tr (X^\dagger X)^{p/2} \right]^{1/p} \ .
\end{equation}
If \eqref{eq:lipschitz-defn} holds, we say $F$ is a $\kappa$-Lipschitz function.
This sort of bound on the behavior of $F$ allows the Gaussian measure \eqref{eq:C-distribution} to reliably determine the probability of fluctuations in $F$.
If $F$ was allowed to vary too wildly in small volumes of the metric \eqref{eq:C-metric}, the Gaussian measure \eqref{eq:C-distribution} could support large fluctuations of $F$ within the high probability regions of the underlying space, and no reasonable bound on such fluctuations would be possible.

We begin by noticing that the Cauchy-Schwarz inequality for the Hilbert space inner product on $\cH_B \otimes \cH_R$ implies
\begin{equation}
    \left( \| X_1 \ket{\psi} \| - \| X_2 \ket{\psi} \| \right)^2 \leq \| (X_1-X_2)\ket{\psi} \|^2\ .
\label{eq:cauchy-schwarz}
\end{equation}
As such, we have
\begin{equation}
\begin{split}
    |F(C_1) - F(C_2)| & \leq F(C_1 - C_2) \\
    & = |B|^{-1/2} \sqrt{\bra{\psi} (C_1 - C_2)^\dagger (C_1-C_2)  \otimes I_R \ket{\psi} } \\
    & \leq |B|^{-1/2} \|(C_1-C_2)\otimes I_R \|_\infty \\
    & = |B|^{-1/2} \| C_1-C_2\|_\infty \\
    & \leq |B|^{-1/2} \| C_1-C_2\|_2\ ,
\end{split}
\end{equation}
where we have defined the spectral norm
\begin{equation}
    \| X \|_\infty \equiv \sup_{\|\ket{\psi}\|=1} \| X\ket{\psi} \|\ ,
\end{equation}
and in the first line we used \eqref{eq:cauchy-schwarz}, in the fourth line we used the spectral norm property $\| X \otimes I \|_\infty = \| X \|_\infty$, and in the last line we used the matrix norm inequality $\|X\|_\infty \leq \|X\|_2$.
Comparing this calculation with \eqref{eq:lipschitz-defn}, we find that $F(C)$ has a Lipschitz constant
\begin{equation}
    \kappa = |B|^{-1/2}\ .
\label{eq:lipschitz-F}
\end{equation}
This constant is quite small if $|B|$ is large, which means $F$ varies only a little bit even over large distances in the metric \eqref{eq:C-metric}.

With the Lipschitz constant \eqref{eq:lipschitz-F} in hand, we can use the deviation bound for the probability distribution \eqref{eq:C-distribution} and metric \eqref{eq:C-metric} which states that any $\kappa$-Lipschitz function $G(C)$ obeys for $\epsilon \in \bR$ (see equations (B.4) and (B.7) in \cite{Akers:2022qdl}):
\begin{equation}
    \Pr [G-\avg{G} \geq \epsilon] \leq \exp \left( -\frac{\epsilon^2}{\kappa^2} \right)\ ,
\label{eq:deviation-bound-G}
\end{equation}
where we have written $\avg{G}$ for the average of $G$ in the distribution \eqref{eq:C-distribution}.
The intuition for this theorem is similar to what we described under \eqref{eq:lipschitz-defn}.
When a function $G$ varies slowly enough, its fluctuations must occur on rather large distance scales in the metric \eqref{eq:C-metric}.
Because the probability measure \eqref{eq:C-distribution} is strongly concentrated in a small region, the high probability region of the underlying space is relatively small compared to the scale on which $G$ fluctuates.
As such, the average value $\avg{G}$ in the measure \eqref{eq:C-distribution} is almost surely within some small distance from the value of $G$ itself as measured with the probability distribution on the underlying space, since the large fluctuations are forced to regions of low probability.\footnote{For a more rigorous treatment, see Appendix B of \cite{Akers:2022qdl}.}

To apply this theorem and prove \eqref{eq:dev-bound-F}, we need to know the average $\avg{F}$ given by
\begin{equation}
    \avg{F} = \int \D C\; F(C)\ .
\end{equation}
Due to the square root in the definition of $F(C)$, it is difficult to evaluate $\avg{F}$ exactly, so we instead bound it above and below.
To bound $\avg{F}$ from above, we use Jensen's inequality
\begin{equation}
    \int \D C\; F(C) \leq \sqrt{\int \D C\; F^2(C) }\ .
\end{equation}
A short calculation shows
\begin{equation}
    |B|^{-1} \int \D C\; C^\dagger C = I_b\ ,
\end{equation}
which is nothing but the statement that $V$ is an isometry on average.
So, we conclude
\begin{equation}
    \avg{F} \leq 1\ .
\end{equation}
To bound $\avg{F}$ from below, we use the fact that for any $x \geq 0$ we have $x \geq \frac{3}{2}x^2 - \frac{1}{2}x^4$, so
\begin{equation}
    \int \D C \; F(C) \geq \int \D C \left( \frac{3}{2}F^2(C) - \frac{1}{2} F^4(C) \right)\ .
\end{equation}
Another short calculation shows
\begin{equation}
    \int \D C \; F^4(C) = 1 + |B|^{-1} \Tr  \psi_b^2\ ,
\end{equation}
where $\psi_b$ is the reduced density matrix
\begin{equation}
    \psi_b \equiv \Tr_R \ket{\psi}\bra{\psi}\ .
\end{equation}
Therefore, we have the lower bound
\begin{equation}
    \avg{F} \geq 1 - \frac{1}{2}|B|^{-1} \Tr \psi_b^2 \equiv 1 - \zeta\ ,
\end{equation}
where we have defined the constant $\zeta$ which obeys
\begin{equation}
    2\zeta \leq |B|^{-1}\ ,
\end{equation}
which follows from the fact that $\psi_b$ is a density matrix and therefore $\Tr \psi_b^2 \leq \Tr \psi_b = 1$.

Applying the deviation bound \eqref{eq:deviation-bound-G} for $F$ from above and below with \eqref{eq:lipschitz-F}, we find (for $\epsilon,\delta > 0$)
\begin{equation}
    \begin{split}
        \Pr[F \geq 1+\epsilon] & \leq \Pr[F-\avg{F} \geq \epsilon] \leq \exp \left(-\epsilon^2 |B| \right)\ ,\\
        \Pr[F \leq 1-\zeta -\delta] & \leq \Pr[\avg{F}-F \geq \delta] \leq \exp \left(-\delta^2 |B| \right)\ .
    \end{split}
\label{eq:deviation-pair}
\end{equation}
Of course, the second of each of these inequalities follows immediately from \eqref{eq:deviation-bound-G} and \eqref{eq:lipschitz-F}.
The nontrivial content here is our use of the bounds on $\avg{F}$ to conclude that the probability of deviations above (below) the particular value 1 ($1-\zeta$) is also strongly bounded from above for large $|B|$.
The first of each of these inequalities follows because a fluctuation in $F$ of size $\epsilon$ ($\delta$) above (below) the mean value $\avg{F}$ is at least as likely as a similar fluctuation above (below) the value 1 ($1-\zeta$), since $\avg{F} \leq 1$ ($\avg{F} \geq 1-\zeta$).

We now choose the parametrization
\begin{equation}
    \epsilon = |B|^{-\gamma}\ , \quad \delta = |B|^{-\gamma} -\zeta\ ,
\end{equation}
which is positive for $\gamma < 1$, and then restrict to $\gamma < 1/2$ and $|B| \geq 4$, which further allows us to write
\begin{equation}
    1 - \zeta |B|^\gamma \geq 1 - \frac{1}{2}|B|^{\gamma-1} \geq 1-\frac{1}{2}|B|^{-1/2} \geq \frac{1}{\sqrt{2}} \ . 
\end{equation}
The purpose of these last few steps is to symmetrize the inequalities, and amounts to the weakening $\delta^2 \to |B|^{-2\gamma}/2$ in the lower deviation bound of \eqref{eq:deviation-pair}.
We can modify the upper deviation bound in \eqref{eq:deviation-pair} by simply replacing $\epsilon^2 \to |B|^{-2\gamma}/2$, as this leaves the inequality valid but weakened.
Finally, we obtain by the union bound $\Pr[A \cup B] \leq \Pr[A]+\Pr[B]$:
\begin{equation}
    \Pr[|F-1| \geq |B|^{-\gamma}] \leq 
    2\exp \left( -\frac{|B|^{1-2\gamma}}{2} \right)\ ,
\end{equation}
as promised in \eqref{eq:dev-bound-F}.

The general arguments in Appendix C of \cite{Akers:2022qdl} can now be used to translate this bound on norm deviation into a bound on the change in overlap of two semiclassical states $\ket{\psi_1}$ and $\ket{\psi_2}$ under the microcanonical dictionary $V$.
Roughly speaking, preservation of norms can be translated to preservation of overlaps at the cost of an $O(1)$ factor by the union bound and an $O(1)$ factor in the preservation accuracy, and another union bound is used to compute the deviation probability of the maximum overlap change in a set $S$ of states.
Specifically, the probability that the largest change in overlap between any two states in a state set $S$ of size $N_S$ is more than an exponentially small (in $\log |B|$) quantity obeys
\begin{equation}
    \Pr \left[ \max_{\ket{\psi_1},\ket{\psi_2} \in S} \Big| \bra{\psi_1} V^\dagger V \otimes I_R \ket{\psi_2} - \inner{\psi_1}{\psi_2} \Big| \geq \sqrt{18} |B|^{-\gamma} \right] \leq 12 \binom{N_S}{2} \exp \left( -\frac{|B|^{1-2\gamma}}{2} \right)\ ,
\label{eq:overlap-bound}
\end{equation}
for $0 < \gamma < 1/2$ and $|B| \geq 4$.

From this result, we see that the typical non-isometric code in the ensemble \eqref{eq:dictionary-3} is expected to preserve the overlaps of a rather large set of states, in fact an even larger set than in \cite{Akers:2022qdl} due to differences between the Haar and complex Gaussian deviation bounds.
These differences conspire to turn the 2 in the denominator of the exponential in \eqref{eq:overlap-bound} into a 24 in the Haar case.
We may reasonably choose $N_S$ to be subexponentially large (in $|B|$), namely $N_S \leq e^{|B|^\nu}$ for $\nu < 1-2\gamma$, and a typical $V$ will still approximately preserve all overlaps with high probability and exponential precision (in $\log |B|$) for large enough $|B|$.
Even the regime of validity has been slightly extended compared to the analogous expression in \cite{Akers:2022qdl}, which held for $|B| \geq 16$ instead.

\subsection{On a relation to complexity theory}\label{sec:relation}

We arrived at \eqref{eq:overlap-bound} by employing the measure concentration formula \eqref{eq:deviation-bound-G}, and this result implies that the pairwise overlaps in a large set of semiclassical states are preserved by a typical dictionary $V$ to exponential precision (in $\log |B|$) with high probability.
However, in deriving \eqref{eq:overlap-bound}, we have not needed to give any details about the set $S$ itself.
This means \eqref{eq:overlap-bound} holds for any $S$ we like, and we may make a convenient choice for it.
In \cite{Akers:2022qdl}, a choice for this set was made to facilitate a relation to complexity theory.
This is certainly quite reasonable from a physical perspective in view of older relations between complexity theory and black hole physics \cite{Harlow:2013tf}.
However, the sense in which this choice exists is rather abstract, and its ambiguity is a result of using averaging to study ``typical'' non-isometric codes of some sort.
To understand this issue, we need to discuss the interpretation of \eqref{eq:overlap-bound} in more detail.

For a specific code, the set of states $S$ with approximately preserved overlaps is relatively unambiguous, up to the existence of more than one such set of maximal size, or perhaps multiple disjoint or overlapping such sets.
So for a specific code, there is no meaning to the probability in \eqref{eq:overlap-bound}, and for any given set $S$ the code either preserves all pairwise overlaps with some precision or it does not.
When we introduce the ensemble of dictionaries and write \eqref{eq:overlap-bound}, we are bounding the probability in the measure \eqref{eq:C-distribution} that a generic set of size $N_S$ will have its pairwise overlaps preserved.
Importantly, this does not mean that we expect overlaps to be preserved in a specific ``typical'' code for arbitrary sets of states as long as they have small enough size as dictated by \eqref{eq:overlap-bound}.
It also does not mean that, for a specific set $S$ with small enough size, we are guaranteed approximate overlap preservation for all codes by \eqref{eq:overlap-bound}.

To understand the meaning of \eqref{eq:overlap-bound}, it is helpful to belabor some elementary facts concerning probability theory.
For some inequality $A(C)$ involving a random variable $C$ with probability distribution $\D C$, the expression $\Pr [A]$ is defined by
\begin{equation}
    \Pr [A] \equiv \int \D C\; \theta_A(C)\ ,
\label{eq:prob}
\end{equation}
where $\theta_A(C)$ is the indicator function
\begin{equation}
    \theta_A(C) = 
    \begin{cases}
        1\ ,\quad A(C) \text{ is true},\\
        0\ ,\quad A(C) \text{ is false}.
    \end{cases}
\end{equation}
The validity of the integration in \eqref{eq:prob} hinges on whether or not the set $\theta_A(C) = 1$ is measurable under $\D C$.
If it is, we call the indicator measurable.

Let $P(C)$ be the inequality appearing in the square brackets in \eqref{eq:overlap-bound} and let $\theta_{P,S}(C)$ be its indicator for the specific set $S$.
In this language, \eqref{eq:overlap-bound} means that for any choice of a specific set $S$ that has $N_S \leq e^{|B|^\nu}$ for $\nu < 1-2\gamma$, the indicator function $\theta_{P,S}$ is equal to zero in a large (as measured by the probability distribution) region in the space of codes.
This large region may be very complicated to describe.
If we change the specific set $S$ to some other $S'$, the indicator functions $\theta_{P,S}$ and $\theta_{P,S'}$ may differ drastically, but their total integrals over the space of codes with measure \eqref{eq:C-distribution} both yield a result that is bounded by the right hand side of \eqref{eq:overlap-bound}, so there must still exist a large region where the new indicator function $\theta_{P,S'}$ is zero.

So the meaning of \eqref{eq:overlap-bound} is that no matter what specific set $S$ we decide to pick, as long as it is small enough in the sense we described and leads to a measurable indicator, it will not be too difficult to find a specific code which preserves pairwise overlaps in $S$ to the accuracy written in \eqref{eq:overlap-bound} when $|B|$ is large.
Unfortunately, \eqref{eq:overlap-bound} tells us nothing about the set of specific codes that do the job for a specific set $S$, so \eqref{eq:overlap-bound} should be thought of as a proof of existence (perhaps more accurately, a proof of abundance) for specific codes preserving overlaps in a specific set $S$ of the requisite size.

The choice made in \cite{Akers:2022qdl} was the set of quantum states in $\cH_b \otimes \cH_R$ with subexponential quantum circuit complexity in the parameter $\log |B|$.\footnote{See \cite{Susskind:2018pmk} for basic definitions and ideas in the quantum complexity of states and operators.}
This amounts to the hypothesis that choosing the set of subexponential states for $S$ leads to a measurable indicator function under \eqref{eq:C-distribution}.
This seems quite reasonable, as the inequalities entering in the definition of the indicator are quite mundane.
The other necessary condition for this choice to be valid was already checked in \cite{Akers:2022qdl}: the size of the set of states with subexponential complexity (relative to a fixed state) grows slowly enough\footnote{There is also a restriction on the size of $|b|$ relative to $|B|$.  Namely, $|b|$ should not be doubly exponential in $|B|$.} as a function of $|B|$ that a deviation bound of the form \eqref{eq:overlap-bound} is strong enough to imply the approximate preservation of overlaps for a set of states of equivalent size with high probability.
However, for a gravitational theory with a fixed dictionary, we cannot appeal to such probabilistic arguments to demonstrate the typicality of overlap preservation for subexponential states.
As such, until the dictionary can be analyzed for a fixed theory, the relation of gravitational non-isometric codes to complexity remains unclear.

An additional subtlety in this complexity hypothesis is the following: the quantum complexity of states $\cC_{\ket{\Omega}}(\ket{\psi})$, unlike that of operators, requires a choice of reference state $\ket{\Omega}$ with respect to which complexity will be measured.\footnote{The complexity of unitary operators does not suffer from such an ambiguity because the identity operator is a preferred element in the group of operators, but the analogous element of a Hilbert space is the zero vector which cannot be used to build nontrivial states by acting with linear operators.}
The non-isometric codes studied in \cite{Akers:2022qdl} came from the Haar ensemble, which is invariant under left and right unitary action.
As such, technically speaking, the results in \cite{Akers:2022qdl} hold for any choice of reference state.
A simple state in some basis may be chosen by hand, but this is an additional input when using the Haar ensemble.
From the gravitational standpoint of this work, we have instead considered an ensemble which has a preferred pair of bases.
These preferred bases may help to resolve the ambiguity in the reference state, as we discuss further in Section~\ref{sec:disc}.

We proceed to discuss bulk reconstruction in  Section~\ref{sec:reconstruction}.
As for the nature of the sets with preserved overlaps for a typical code in \eqref{eq:dictionary-3}, we will not make a choice for them in this work.
Instead, we will consider an abstract set $S$ of states which has all pairwise overlaps preserved in a large region of the probability space as in \eqref{eq:overlap-bound}.
We are more or less guaranteed the existence of specific codes which preserve the overlaps in any such $S$ if it has subexponential (in $|B|$) size due to \eqref{eq:overlap-bound}.
We will see that simply having such a set is sufficient to get an interesting reconstruction theory, and the particular set in question (whether it consists of subexponential states, bounded energy states, or any other characterization) is not so important for the analysis.
We refer to states in this set, whatever it may be, as $S$-states.

\section{Bulk reconstruction}\label{sec:reconstruction}

We begin by recalling the idea of unitary operator reconstruction in standard isometric error correction.
Reconstruction of a code space unitary operator $W$ in the case of an isometric code $V$ means producing a physical unitary operator $\tW$ that satisfies
\begin{equation}
    \| \tW V \ket{\psi} - V W \ket{\psi} \| \leq \epsilon\ ,
\label{eq:basic-recon}
\end{equation}
for some small error parameter $\epsilon$ and any $\ket{\psi} \in \cH_b \otimes \cH_R$.
When $V$ is isometric, we may write a generic formula $\tW = VWV^\dagger$ which is an exact reconstruction on the total physical Hilbert space with zero error.

If we wish to reconstruct a code space operator with a physical operator that acts only on $\cH_B$ or $\cH_R$, the situation can be more complicated depending on the structure of the states in the code space, and we may only be able to reconstruct $W$ approximately, if at all.
However, as long as $V$ is isometric, we may still use a formula rather similar to $VWV^\dagger$ which is known as the Petz map \cite{Petz:1986tvy,Petz:1988usv}.
For instance, we can try to reconstruct a code unitary $W$ using a physical unitary $\tW_B$ which acts only on $\cH_B$.
In this case, the Petz map formula for $\tW_B$ is
\begin{equation}
    \tW_B = \Tr_R^{-1/2} (VV^\dagger) \Tr_R (V W V^\dagger)  \Tr_R^{-1/2} (VV^\dagger)\ .
\label{eq:petz-example}
\end{equation}

Depending on the structure of the states in our code space, the Petz reconstruction formula \eqref{eq:petz-example} may or may not work.
If $V$ is an isometry and approximate reconstruction is possible on the specified physical subspace, it is guaranteed to work reasonably well, at least for a typical code state.
But, if $V$ is highly non-isometric, it might not succeed even if reconstruction is technically possible in the sense that there exists a physical unitary operator which implements \eqref{eq:basic-recon} for the code state of interest.
In fact, when $V$ is non-isometric, not even the state-independent global reconstruction $VWV^\dagger$ is guaranteed to succeed because $\| V^\dagger V \ket{\psi} - \ket{\psi} \|$ can be very large.

Indeed, from the expression \eqref{eq:basic-recon}, it is clear that a map $V$ which does not preserve the norm of a state $W\ket{\psi}$ but does preserve the norm of $\ket{\psi}$ cannot support reconstruction of $W$.
This is because $W\ket{\psi}$ and $V\ket{\psi}$ will have roughly unit norm, but then no action of a physical unitary operator $\tW$ on $V\ket{\psi}$ can reproduce the norm-changing action of $V$ on $W\ket{\psi}$, making \eqref{eq:basic-recon} impossible to satisfy.
This is quite a different situation than in isometric codes, where reconstruction of any unitary acting on any state in the code space is always possible by the global reconstruction $VWV^\dagger$.
It is furthermore clear that any operator $W$ for which $V$ preserves the norms of all $W \ket{\psi}$ for all $S$-states can be reconstructed on all $S$-states as long as we allow the operator $\tW$ to depend on the input state $\ket{\psi}$.
If two states $V\ket{\psi}$ and $VW\ket{\psi}$ have approximately the same norm, they are approximately related by a unitary rotation in $\cH_B \otimes \cH_R$ which we may take as $\tW(\psi)$.
This state dependence is in fact necessary for non-isometric reconstruction \cite{Akers:2022qdl}.

From what we have just described, we conclude that the simplest set of unitary operators reconstructible on all of $S$ for a typical non-isometric code of the form \eqref{eq:dictionary-3} is the one which preserves the set $S$ itself, which we recall is a set of states which have approximately preserved overlaps under the action of $V$ as in \eqref{eq:overlap-bound}.
We will refer to such unitary operators as $S$-operators.
If we are not interested in reconstructing $W$ on every $S$-state, we need not require that $W$ preserve the entire set $S$, but rather we only need that $W$ maps the state in which we are interested (in $S$) to another state in $S$.\footnote{We have ignored the possibility of an ``accidental'' reconstruction where $V$ happens to change the norm of $W\ket{\psi}$ and $\ket{\psi}$ in precisely the same way.  It would be interesting to understand when and how this can occur in non-isometric codes, but we will not do so here.}
But, the code operators we consider most interesting are those which may be reconstructed in a state-specific manner on any state in $S$, so we choose preservation of $S$ as our definition of an $S$-operator.
We now recall the formalization of state-specific reconstruction following \cite{Akers:2021fut,Akers:2022qdl}.

\subsection{State-specific reconstruction}

Unfortunately, for non-isometric codes, there is no known analogue of the global reconstruction or Petz map which is guaranteed to succeed when reconstruction is possible \cite{Akers:2022qdl}.
Instead, we must work out the reconstruction formula for a generic code unitary $W$ by using the explicit structure of the dictionary $V$.
Moreover, the resulting reconstructed unitary operator $\tW$ will necessarily depend on the code state $\ket{\psi}$ upon which the code operator $W$ is intended to act \cite{Akers:2022qdl}.
So, by a state-specific reconstruction on e.g.~$\cH_B$ for an encoding map $L$, we mean an operator $\tW_B(\psi)$ which satisfies
\begin{equation}
    \| \tW_B(\psi) L\ket{\psi} - LW\ket{\psi} \| \leq \epsilon\ .
\label{eq:state-specific}
\end{equation}

Before trying to reconstruct explicit operators, it is helpful to understand when a reconstruction is expected to exist; this task will be our main focus.
In \cite{Akers:2022qdl}, it was shown that an operator $\tW_B(\psi)$ exists satisfying \eqref{eq:state-specific} for the encoding map $L$ if the decoupling bound
\begin{equation}
    \| \Tr_B (LW\ket{\psi}\bra{\psi}W^\dagger L^\dagger) - \Tr_B (L\ket{\psi}\bra{\psi}L^\dagger) \|_1 \leq \epsilon^2\ ,
\label{eq:non-iso-decoupling}
\end{equation}
is obeyed, where the $\epsilon$ appearing in \eqref{eq:non-iso-decoupling} is the same as the one in \eqref{eq:state-specific}.
Furthermore, with $L = V \otimes I_R$, for $S$-states and $S$-operators \eqref{eq:state-specific} also implies \eqref{eq:non-iso-decoupling} with a slightly different relationship between the infinitesimal parameters, so \eqref{eq:non-iso-decoupling} and \eqref{eq:state-specific} are effectively equivalent for these states and operators.

Following the arguments in Sections 3 and 5 of \cite{Akers:2022qdl}, we first note that state-specific physical unitary reconstructions $\tW_{BR}(\psi)$ exist for all $S$-states $\ket{\psi}$ and $S$-operators $W$ due to the triangle inequality along with \eqref{eq:overlap-bound}.
The relevant decoupling bound to check is \eqref{eq:non-iso-decoupling} with traces over $\cH_B \otimes \cH_R$ instead of only $\cH_B$, and this gives
\begin{equation}
    \begin{split}
        \big| \bra{\psi} W^\dagger (V^\dagger V \otimes I_R) W \ket{\psi} - \bra{\psi} V^\dagger V \otimes I_R \ket{\psi} \big| & = \Big| \left( \bra{\psi} W^\dagger (V^\dagger V \otimes I_R) W \ket{\psi} - \inner{\psi}{\psi} \right) \\
        & \quad - \left( \bra{\psi} V^\dagger V \otimes I_R \ket{\psi} - \inner{\psi}{\psi} \right) \Big| \\
        & \leq \big| \bra{\psi} W^\dagger (V^\dagger V \otimes I_R) W \ket{\psi} - \bra{\psi} W^\dagger W \ket{\psi} \big| \\
        & \quad + \big| \bra{\psi} V^\dagger V \otimes I_R \ket{\psi} - \inner{\psi}{\psi} \big| \\
        & \leq 2 \sqrt{18} |B|^{-\gamma}\ ,
    \end{split}
\label{eq:global-state-spec-recon}
\end{equation}
where in the second line we used the unitary property $W^\dagger W = I$ and the triangle inequality $|a+b| \leq |a| + |b|$.
In the third line we simply applied \eqref{eq:overlap-bound} since $\ket{\psi}$ is an $S$-state and $W$ is an $S$-operator so $W\ket{\psi}$ is also an $S$-state.
We stress that this result holds for all $S$-operators without qualification, since it only makes use of the property in \eqref{eq:overlap-bound} which holds for all $S$-states.

Furthermore, if the code $S$-operator acts only on $\cH_R$ as $W_R$, the trivial state-independent reconstruction $\tW_R = W_R$ will work for all $S$-states.
In fact, for all operators $W_R$ (not just $S$-operators), we have the commutation $\tW_R (V \otimes I_R) = (V \otimes I_R) W_R$, so \eqref{eq:basic-recon} holds exactly for all states (not just $S$-states) with the trivial reconstruction.
The restriction to $S$-states and $S$-operators arises if we also wish to require that matrix elements of $\tW_R$ in the physical space images of code states should match matrix elements of $W_R$ in those code states.
This can be ensured by writing
\begin{equation}
\begin{split}
    \bra{\psi_2} (V^\dagger \otimes I_R) \tW_R (V \otimes I_R) \ket{\psi_1} & = \bra{\psi_2} V^\dagger V \otimes W_R \ket{\psi_1} \\
    & \approx \bra{\psi_2} W_R \ket{\psi_1}\ ,
\end{split}
\label{eq:trivial-R-recon}
\end{equation}
where in the first line we used $\tW_R = W_R$ and in the second line we used the fact that $\ket{\psi_1}$ and $\ket{\psi_2}$ are $S$-states, $W_R$ is an $S$-operator so $W_R\ket{\psi_1}$ is an $S$-state, and then by \eqref{eq:overlap-bound} we have approximate overlap preservation for all $S$-states.
This secondary condition is implied by \eqref{eq:basic-recon} for an isometric $V$.
But as we have just seen, that relationship fails if $V$ is non-isometric.
We might have \eqref{eq:basic-recon} for some states and operators on which \eqref{eq:trivial-R-recon} could fail, and such failures can occur precisely when the overlap between $W_R\ket{\psi_1}$ and $\ket{\psi_2}$ is not preserved by $V$.
Indeed, restricting to $S$-states and $S$-operators allows us to once again derive the analogous secondary condition from \eqref{eq:state-specific} for a state-specific reconstruction $\tW(\psi_1)$ of a code operator $W$:
\begin{equation}
\begin{split}
    \bra{\psi_2} (V^\dagger \otimes I_R) \tW(\psi_1) (V \otimes I_R) \ket{\psi_1} & \approx \bra{\psi_2} (V^\dagger V \otimes I_R) W \ket{\psi_1} \\
    & \approx \bra{\psi_2} W \ket{\psi_1}\ ,
\end{split}
\end{equation}
where in the first line we used \eqref{eq:state-specific} and in the second line we used \eqref{eq:overlap-bound} and the restriction to $S$-states $\ket{\psi_1}$ and $\ket{\psi_2}$ and $S$-operators $W$.

The case where the code operator acts only on $\cH_b$ and ought to be reconstructed in only $\cH_B$ or $\cH_R$ is harder. 
As usual, the global reconstruction $\tW_B = VW_bV^\dagger$ will fail when $V$ is highly non-isometric.
But this failure cannot be remedied by simply restricting to $S$-states and $S$-operators as we did for the $\tW_{BR}(\psi)$ or $\tW_R$ reconstructions in \eqref{eq:global-state-spec-recon} and \eqref{eq:trivial-R-recon}, because \eqref{eq:state-specific} cannot be written as a difference between two overlaps that only differ by an insertion of $V^\dagger V \otimes I_R$.
So we cannot use \eqref{eq:overlap-bound} to argue that the global reconstruction will hold for general $S$-states and $S$-operators.
Instead, we must proceed by deriving another deviation bound similar to \eqref{eq:dev-bound-F}, but this time for the decoupling criterion \eqref{eq:non-iso-decoupling}.

\subsection{Entanglement wedge reconstruction}

The only case which remains is when the operator of interest $W_b$ acts only on $\cH_b$, and we wish to reconstruct it with either $\tW_B(\psi)$ or $\tW_R(\psi)$.
This will be possible in situations roughly determined by entanglement wedge reconstruction. 
When the interior is in the entanglement wedge of $B$, we will be able to reconstruct $W_b$ with $\tW_B(\psi)$ on $\cH_B$, and similarly for reconstruction on $\cH_R$.

When $W_b$ is an operator that we expect to reconstruct on $\cH_B$, we must use the decoupling principle \eqref{eq:non-iso-decoupling} to demonstrate the existence of $\tW_B(\psi)$.
The relevant quantity to study is
\begin{equation}
    \Psi_{BR}(V,W) = (V \otimes I_R) W \ket{\psi}\bra{\psi} W^\dagger (V^\dagger \otimes I_R)\ ,
\end{equation}
and demonstrating the decoupling bound amounts to verifying the relation
\begin{equation}
    \| \Psi_R(V,W_b) - \Psi_R(V,I) \|_1 \leq \epsilon^2\ ,
\end{equation}
for some small $\epsilon$, where $\Psi_R = \Tr_B \Psi_{BR}$.

We begin by demonstrating that this bound holds on average for a particular generic $W_b$.
We use the $p$-norm bound $\| X_A \|_1 \leq \sqrt{|A|} \|X_A\|_2$ for operators $X_A: \cH_A \to \cH_A$ followed by Jensen's inequality to write
\begin{equation}
    \int \D C\; \| \Psi_R(V,W_b) - \Psi_R(V,I) \|_1 \leq \sqrt{ |R| \int \D C\; \| \Psi_R(V,W_b) - \Psi_R(V,I_{bR}) \|_2^2 }\ .
\end{equation}
Evaluating the integral, we find
\begin{equation}
    \int \D C\; \| \Psi_R(V,W_b) - \Psi_R(V,I) \|_2^2 = \frac{2}{|B|} \left[ 1 - |\bra{\psi} W_b \ket{\psi}|^2 \right]\ ,
\end{equation}
which, due to the inequality $|\bra{\psi} W_b \ket{\psi}|^2 \geq 0$, implies the bound
\begin{equation}
    \int \D C\; \| \Psi_R(V,W_b) - \Psi_R(V,I) \|_1 \leq \sqrt{\frac{2|R|}{|B|}}\ .
\label{eq:B-recon-L1-bound}
\end{equation}
A similar result holds for reconstruction on the $R$ system with
\begin{equation}
    \int \D C\; \| \Psi_B(V,W_b) - \Psi_B(V,I) \|_2^2 = 2 \left[ \Tr \psi_b^2 - \Tr (W_b \psi_b W_b^\dagger \psi_b ) \right]\ ,
\end{equation}
implying the bound
\begin{equation}
    \int \D C\; \| \Psi_B(V,W_b) - \Psi_B(V,I) \|_1 \leq \sqrt{2|B| \Tr \psi_b^2 }\ ,
\label{eq:R-recon-L1-bound}
\end{equation}
because by evaluating the trace in the eigenbasis of $\psi_b$ we conclude that the expression $\Tr (W_b\psi_bW_b^\dagger\psi_b)$ is bounded below by the smallest eigenvalue of $\psi_b$, which is non-negative since $\psi_b$ is a density matrix.

These results are consistent with entanglement wedge reconstruction in the following sense.
When $|R| \ll |B|$, the right hand side of \eqref{eq:B-recon-L1-bound} is small and we have an accurate state-specific reconstruction $\tW_B(\psi)$ by the decoupling theorem \eqref{eq:non-iso-decoupling}.
On the other hand, when the second bulk R\'enyi entropy $S^{(2)}_\psi(b)$ is much larger than $\log |B|$, the right hand side of \eqref{eq:R-recon-L1-bound} is small since $\Tr \psi_b^2 = \exp(-S^{(2)}_\psi(b))$, and we have an accurate state-specific reconstruction $\tW_R(\psi)$ by the analogous versions of \eqref{eq:state-specific} and \eqref{eq:non-iso-decoupling}.

Entanglement wedge reconstruction in our context simply says that when the reservoir von Neumann entropy obeys $S_\psi(R) \ll \log |B|$, $W_b$ should be reconstructible as $\tW_B(\psi)$, and when $S_\psi(R) \gg \log |B|$, $W_b$ should be reconstructible as $\tW_R(\psi)$.
The inequalities $|R| \ll |B|$ and $S^{(2)}_\psi (b) \gg \log |B|$ indeed imply these expressions, respectively.
This is because the von Neumann entropy obeys $S_\psi (R) \leq \log |R|$, so $|R| \ll |B|$ implies $S_\psi(R) \ll \log |B|$.
Furthermore, when $S^{(2)}_\psi (b) \gg \log |B|$, we have $S^{(2)}_\psi (R) \gg \log |B|$ since pure states obey $S^{(2)}_\psi (b) = S^{(2)}_\psi (R)$, and then the inequality $S_\psi (R) \geq S^{(2)}_\psi (R)$ implies $S_\psi (R) \gg \log |B|$.
Thus, our state-specific reconstructions exist within the regimes predicted by entanglement wedge reconstruction.

Note that these reconstruction results apply very generally due to the state-specificity. 
We can give reconstructions of $W_b$ for states which have very different amounts of bulk entropy, from a zero entropy product state between $\cH_b \otimes \cH_R$ to a semiclassical state that naively has ``too much'' entropy with $S_\psi (b) \gg \log |B|$ and runs into the Hawking paradox.
All such states, as long as they are $S$-states, are treated on an equal footing in the code space.

To argue that the decoupling bound \eqref{eq:non-iso-decoupling} holds for a larger set of $S$-operators $W_b$ and $S$-states $\ket{\psi}$ and not just particular such operators and states, we need another deviation bound.
Following Appendix~F of \cite{Akers:2022qdl}, we will bound deviations of
\begin{equation}
    K_R(C) \equiv \frac{ \| \Psi_R(V,W_b) - \Psi_R(V,I) \|_1}{\| (V \otimes I_R)W_b\ket{\psi} \| + \| (V \otimes I_R) \ket{\psi} \|  }\ ,
\label{eq:K-defn}
\end{equation}
in the case of reconstruction on $B$, and a similar expression $K_B(C)$, using $\Psi_B$ instead of $\Psi_R$, in the case of reconstruction on $R$.
To obtain a Lipschitz constant for \eqref{eq:K-defn}, we note that the arguments in Appendix~F of \cite{Akers:2022qdl} apply generally enough to give a Lipschitz constant for the right hand side of \eqref{eq:K-defn} for any $V$ simply by knowing the Lipschitz constant for $\| (V \otimes I_R) \ket{\psi} \|$.
For our choice of $V$, we derived the Lipschitz constant \eqref{eq:lipschitz-F}, and therefore by the arguments in \cite{Akers:2022qdl} we have
\begin{equation}
    | K_R(C_1) - K_R(C_2) | \leq 6|B|^{-1/2} \| C_1 - C_2 \|_2\ .
\end{equation}
Combined with \eqref{eq:B-recon-L1-bound} and the complex Gaussian deviation bound \eqref{eq:deviation-bound-G}, we conclude\footnote{Here we are being rather fast.  See Appendix F of \cite{Akers:2022qdl} for more details.}
\begin{equation}
\begin{split}
    \Pr \left[ K_R(C) \geq (2|R|/|B|)^{1/4} + |B|^{-\gamma} \right] & \leq \Pr \left[ K_R(C) \geq \int \D C\; K_R(C) + |B|^{-\gamma} \right] \\
    & \leq \exp \left( -\frac{|B|^{1-2\gamma}}{36} \right)\ ,
\end{split}
\label{eq:B-recon-bound}
\end{equation}
where we have used Jensen's inequality, \eqref{eq:B-recon-L1-bound}, and the fact that $K_R(C)$ is bounded by the denominator in \eqref{eq:K-defn} to conclude $\int \D C \; K_R(C) \leq (2|R|/|B|)^{1/4}$.
A similar bound exists for the quantity $K_B(C)$ associated with reconstruction by $\tW_R(\psi)$:
\begin{equation}
    \begin{split}
    \Pr \left[ K_B(C) \geq (2|B|\Tr \psi_b^2)^{1/4} + |B|^{-\gamma} \right] & \leq \Pr \left[ K_B(C) \geq \int \D C\; K_B(C) + |B|^{-\gamma} \right] \\
    & \leq \exp \left( -\frac{|B|^{1-2\gamma}}{36} \right)\ .
\end{split}
\label{eq:R-recon-bound}
\end{equation}

The utility of deriving \eqref{eq:B-recon-bound} and \eqref{eq:R-recon-bound} is as follows.
As we argued in Section~\ref{sec:overlap}, for $S$-states $\ket{\psi}$ and $S$-operators $W_b$ the denominator of \eqref{eq:K-defn} is highly likely to be approximately 2, and deviations from this value are exponentially (in $|B|$) suppressed in probability.
Therefore, bounding deviations of the function $K_R(C)$ is essentially the same, for $S$-states and $S$-operators, as bounding deviations of the numerator of $K_R(C)$, which is the relevant quantity appearing in the decoupling criterion \eqref{eq:non-iso-decoupling}.
So, these inequalities are providing information similar to \eqref{eq:overlap-bound}: they allow us to estimate, with the union bound, how many $S$-operators may be reconstructed on a given subspace in a state-specific manner.

As we previously estimated the maximum number of $S$-states $N_S$ to be subexponential $N_S \leq e^{|B|^\nu}$ for $\nu < 1-2\gamma$, to apply the union bound and recover a result like \eqref{eq:overlap-bound} we need only choose the number of $S$-operators appropriately.
As long as we consider a set of $S$-operators which is polynomial in the number of $S$-states, namely subexponential in a constant multiple of the black hole Hilbert space dimension, we can guarantee reconstruction of any such $S$-operator $W_b$ as $\tW_R(\psi)$ or $\tW_B(\psi)$ (as appropriate) on any $S$-state with high probability.
The sense in which this choice exists is similar to the choice of $S$ itself: for a fixed code and set $S$, there will be an unambiguous set of operators for which the decoupling theorem holds for subspace reconstruction, and here we are simply estimating how large it will be for a typical code in \eqref{eq:dictionary-3}.
Note that this restriction on $S$-operator reconstruction is only applicable for subspace reconstruction problems.
The global state-specific reconstruction \eqref{eq:global-state-spec-recon} relies only on preservation of $S$.

Now, for large values of $|B|$, the probabilities in \eqref{eq:B-recon-bound} and \eqref{eq:R-recon-bound} are exponentially suppressed.
What this means is that deviations above the mean values of $K_R(C)$ or $K_B(C)$ are highly unlikely.
In the case of e.g.~reconstruction on $B$ in \eqref{eq:B-recon-bound}, the approximate mean value away from which we are bounding deviations is scaling with $|R|/|B|$.
When $|R|/|B|$ is small, this mean value is small, and we are very likely to satisfy the decoupling criterion \eqref{eq:non-iso-decoupling}.
Similar statements hold for reconstruction on $R$, where the relevant mean value is now scaling with $|B| \Tr \psi_b^2$, which is only small when the bulk entropy is much larger than $\log |B|$.
Thus, the average reconstruction results we discussed previously for a particular operator and state can be extended by these deviation bounds to hold for all $S$-states and a relevant subset of $S$-operators in a manner that is consistent with entanglement wedge reconstruction.

Unlike in the case of the overlap bound \eqref{eq:overlap-bound}, we have not studied negative fluctuations from the mean values in \eqref{eq:B-recon-bound} or \eqref{eq:R-recon-bound}.
This is because we were most concerned with verifying that state-specific subspace reconstruction is possible in the regimes that we expect it to be from entanglement wedge reconstruction.
But unexpectedly large negative fluctuations in the variables we analyzed above would only cause reconstruction to be possible in regimes where we would not normally expect it to be from the entanglement wedge.
Such fluctuations would not invalidate our demonstration of reconstruction in the expected regimes.
We could of course explicitly study these negative fluctuations by finding lower bounds for $\avg{K_B(C)}$ and $\avg{K_R(C)}$ to see how sharply entanglement wedge reconstruction controls the exact set of operators which may be reconstructed on subspaces, but we will not do so here.

We note for completeness that there are certain $S$-states and $S$-operators for which reconstruction will not be possible on only $B$ or $R$.
This manifests in the deviation bounds we discussed before as a gap between the regimes where the decoupling criterion is satisfied for either $B$ or $R$.
Specifically, it may be the case that neither $|R|/|B|$ nor $|B|\Tr \psi_b^2$ is small, and the mean values of the quantities appearing in the decoupling criterion are therefore large for both the $B$ and $R$ systems.
In the bulk, this arises from a failure of naive subregion duality for certain states and operators, and this failure can result in $O(1/G_N)$ corrections to the quantum extremal surface prescription \cite{Akers:2020pmf}.\footnote{In understanding the limits of subregion duality more generally, one must consider more fine-grained notions of encoding quantum information along the lines of zero- or alpha-bits \cite{Hayden:2017xed,Hayden:2018khn,Akers:2021fut}.}

Of course, a global state-specific reconstruction $\tW_{BR}(\psi)$ is always possible via \eqref{eq:global-state-spec-recon} for all $S$-states and all $S$-operators if we allow use of the entire physical space $\cH_B \otimes \cH_R$.
This global reconstruction will only fail when our norm preservation results for $V$ fail, and as discussed in Section~\ref{sec:overlap} this can only occur for states which lie outside $S$ or operators which do not preserve $S$.

\section{Discussion}\label{sec:disc}

In this work, we have constructed and studied an ensemble of non-isometric error correcting codes in dilaton gravity.
We argued using measure concentration that the typical code in our ensemble will preserve the norms and pairwise overlaps of a large set of states $S$ in the code space, and estimated its size to be subexponential in the black hole Hilbert space dimension using \eqref{eq:overlap-bound}.
Furthermore, we demonstrated that a state-specific reconstruction of any code unitary operator $W$ which preserves $S$ is possible for any $S$-state, and we found that the support of these $S$-operator reconstructions is consistent with expectations from entanglement wedge reconstruction.

We now turn to remaining questions, several of which are related to complexity theory.
Complexity theory played a role in \cite{Akers:2022qdl}, where it was chosen as a set of $S$-states in the sense we discussed in Section~\ref{sec:relation}.
As we mentioned in Section~\ref{sec:overlap}, we have not attempted to characterize $S$ in any particular way in this work, as such a specific choice is not strictly necessary for studying bulk reconstruction in the ensemble of non-isometric codes we defined.
That being said, it is certainly important to understand whether or not codes defined in specific holographic theories (as opposed to an ensemble) admit sets $S$ which include the set of subexponential states.
As a first step toward this goal, we will point out that our construction has a few interesting features which may be connected to complexity theory, and we discuss these along with related confusions below.
We also address fundamental averaging and the sense in which diffeomorphism invariance and other gravitational properties are present for the typical code we have studied.
Finally, we comment on the breakdown of effective field theory in the bulk.

\subsection{The equilibrium basis}

The ensemble of codes we studied has an interesting property: it has preferred bases in which the individual matrix elements of the dictionary are independent random variables.
We take this as evidence that each individual non-isometric code in our ensemble comes equipped with a preferred basis of code space states and a preferred basis of physical space states.
As we discussed in Section~\ref{sec:dictionary}, the existence of these preferred bases is due to two facts, namely the superselection effect of the brane flavors and the existence of the two-sided fixed asymptotic energy basis for $\cH_\text{grav}$.
We will refer to the preferred code space basis as the ``equilibrium basis'', loosely inspired by older ideas concerning the ``equilibrium state'' which enters in other proposals for bulk reconstruction of the black hole interior \cite{Papadodimas:2012aq}.
There are several differences between the two ideas, some of which are intrinsic to the non-isometric code framework and were addressed already in \cite{Akers:2022qdl}.

The equilibrium basis as we have described it seems rather special to dilaton gravity.
But it is worth trying to understand whether we should expect something like it in more general theories.
In a more realistic theory, it is not clear how this basis would be determined in the bulk.
But, one possibility which deserves more thought is the following.
The equilibrium basis is in some sense created by pushing all excitations deep into the interior, so that the only structure which remains in the state is in the deep infrared.
There has been recent progress in placing constraints on this structure for the zero energy states that source the entropy of supersymmetric black holes \cite{Lin:2022rzw,Lin:2022zxd}, and perhaps a similar bulk picture for general black holes would give us a clue toward the equilibrium basis in general theories.
Also relevant are the results of \cite{Blommaert:2021etf}, which appear to suggest that deep interior dynamics can determine the coefficients of the dictionary in a specific preferred bulk basis.
The rough picture is that the preferred bulk basis is again formed by Euclidean path integral states but with strong bulk interactions localized near some complicated interior structure.

As far as the preferred boundary basis is concerned, there is reason to believe that the energy eigenbasis is not all that important on the boundary in richer theories of gravity.
It is important in dilaton gravity due to the existence of the $\ket{E}_\text{grav}$ basis which forces asymptotic regions connected in the bulk to have the same energy.
But more complicated theories of gravity can support wormholes with different asymptotic energies \cite{Cotler:2021cqa}.
Unfortunately there does not seem to be a candidate to replace the energy eigenbasis, as it is often the only preferred basis that exists in a simple theory of quantum mechanics with only a Hilbert space and Hamiltonian.
So without understanding the detailed structure of the boundary theory, we may not be able to determine the preferred boundary basis.
This is similar to the issue concerning the relevance of interior dynamics for the equilibrium basis.

\subsection{Relative subexponential states}\label{sec:relative}

As complexity was suggested to be relevant for non-isometric codes in \cite{Akers:2022qdl}, the existence of the equilibrium basis suggests a natural conjecture to make contact with complexity: perhaps the equilibrium basis should be thought of as a set of reference states from which to measure complexity.
This is a notion of quantum state complexity relative to multiple states, and to define the complexity $\cC(\ket{\psi}_b)$ of a general state we simply minimize over the relative complexity with any one of the Euclidean path integral bulk states via
\begin{equation}
    \cC(\ket{\psi}_b) \equiv \min_{\alpha}\; \cC_{\ket{\textbf{\br}_\alpha(E)}} (\ket{\psi}_b)\ .
\end{equation}

While this defines complexity for states in $\cH_b$, we must also give a prescription for the complexity of states in $\cH_b \otimes \cH_R$.
As the $R$ system has no dynamics or structure, we cannot hope to select a preferred reference state in $\cH_R$ by itself.
Instead, we fix an arbitrary state $\ket{0} \in \cH_R$ and consider the relative complexity of states in $\cH_b \otimes \cH_R$ with respect to the product states $\ket{\textbf{\br}_\alpha(E)} \otimes \ket{0}_R$ with the Euclidean path integral basis states of $\cH_b$:
\begin{equation}
    \cC(\ket{\psi}_{bR}) \equiv \min_\alpha\; \cC_{\ket{\textbf{\br}_\alpha(E)} \otimes \ket{0}_R} (\ket{\psi}_{bR})\ .
\label{eq:relative-complexity}
\end{equation}
If $R$ is a quantum computer, a reasonable choice for $\ket{0}_R$ is a disentangled product state of qubits.
If instead $R$ is a laboratory outside a black hole where Hawking radiation is to be collected, $\ket{0}_R$ may be the ground state of the detector device.
But, without this sort of extra information about the $R$ system, which will in general depend on the precise physical situation at hand, we cannot say more.

As a first step toward analyzing the implications of our multi-state relative complexity quantity \eqref{eq:relative-complexity}, we need to calculate the number of states with subexponential complexity in $\log |B|$ when complexity is measured by \eqref{eq:relative-complexity} rather than relative to a fixed state as in \cite{Akers:2022qdl}.
This will allow us to check whether the set of subexponential states, with our notion of subexponentiality, can fit within a set of $S$-states which have approximately preserved overlaps.
It is clear that our notion of subexponential will be more restrictive than the one considered in \cite{Akers:2022qdl} since there are more states to use as references.

Generically, we can bound the number $N_\cC$ of $n$-qubit states with complexity $\cC$ by bounding the number of distinct circuits with $\cC$ elementary gates.
If the gate set has size $\Gamma$, this bound is
\begin{equation}
    N_\cC \leq e^{(2\log n + \log \Gamma) \cC}\ .
\end{equation}
If there are $m$ possible initial states that we will consider to have zero complexity, then the corresponding bound on the number of states with complexity $\cC$ can be obtained by simply multiplying the above bound by $m$.
This is admittedly a crude approximation.
It will be a fairly loose bound if there are sizable intersections between the images of subexponential circuits acting on any two reference states.
The number of circuits, and therefore the number of states by \eqref{eq:relative-complexity}, with subexponential complexity $\cC \leq |B|^\upsilon$ is bounded by
\begin{equation}
    N_{\text{subexp}} \leq \exp (\log |b| + (2 \log \log |b| + \log \Gamma) |B|^\upsilon )\ .
\end{equation}
This result means that, for $|b| \leq e^{|B|^\xi}$, we can almost surely guarantee approximate overlap preservation by \eqref{eq:overlap-bound} by choosing $\upsilon$ and $\xi$ such that
\begin{equation}
    \max (\upsilon,\xi) < 1-2\gamma\ ,
\label{eq:regime}
\end{equation}
for $0 < \gamma < 1/2$ and large $|B|$.
From the symmetry between $\upsilon$ and $\xi$ in this expression, it is clear that the multi-state relative complexity \eqref{eq:relative-complexity} equalizes the effects of having a subexponential (in $|B|$) number of reference states with the power to act with a unitary operator of subexponential (in $\log |B|$) complexity.
As such, when \eqref{eq:regime} holds, the set of subexponential states as measured by \eqref{eq:relative-complexity} does indeed fit inside the typical set $S$ with size estimated by \eqref{eq:overlap-bound}.

Note that the Page transition, which is possible to observe in certain highly entangled states in the code space when $|b| \geq |B|$, occurs when $|b|$ is only linear in $|B|$.
So, there is still no issue with reconstructing operators in the interior in this parameter regime.
However, when $|b|$ scales exponentially with $|B|$, we cannot construct a basis for $\cH_b$ where all states have subexponential complexity (in $\log |B|$) relative to a fixed state.
This is because a basis would require at least $e^{|B|}$ states, but the total number of subexponential states relative to a fixed state is asymptotically less than $e^{|B|^\epsilon}$ for any $\epsilon > 0$.
So, the definition \eqref{eq:relative-complexity} becomes truly inequivalent with a standard notion of complexity relative to a single reference state.
By inequivalent in this context we mean that there is an exponential separation between the complexity assigned by \eqref{eq:relative-complexity} and the complexity assigned relative to a single reference state.

\subsection{Limitations on complexity}

So far, we have not provided any concrete evidence for or against the statement that subexponential complexity states and operators are relevant for non-isometric error correction in a fixed theory of gravity.
All of our arguments have been probabilistic and independent of complexity theory.
Instead of addressing this issue directly, we will now discuss some general limitations on how complexity may enter the non-isometric structure for a fixed dictionary $V$.

The most naive idea for how complexity may be relevant is that, while overlaps of pairs of states in the equilibrium basis are preserved with very high accuracy by the dictionary $V$ in the regime \eqref{eq:regime}, acting with simple unitary operators on $\cH_b$ should take us outside the equilibrium basis and slowly change the norm of the state.
With each application of a simple unitary operator (perhaps an element of some elementary gate set), the norm of the particular bulk state will change slightly under the dictionary $V$.
This effect will be negligible until an exponential (in $\log |B|$) number of elementary gates have been applied, at which point there will be a large change in the norm as \eqref{eq:overlap-bound} can no longer guarantee with good probability that the norm is preserved for such a large class of states.
In this way, the norm of the microscopic state is meant to differentiate between subexponential and exponential complexity states in the semiclassical Hilbert space.
This is relevant for the reconstruction problem because when the norm of the state under $V$ is greatly modified, there is no way to act unitarily on the microscopic Hilbert space in a manner that reproduces this modification, as we mentioned in Section~\ref{sec:reconstruction}.

In light of certain no-go results in complexity theory concerning the general problem of determining the algorithmic complexity of a given state or operator \cite{Kolmogorov:1968thr,ChaitinAC95,RAZBOROV199724}, we might be skeptical that the norm of the image under $V$ of a state can really detect something like the complexity of the state, even at the most coarse-grained level of determining whether or not the complexity is exponential.
Indeed, because the quantum circuit complexity of states is essentially a notion of algorithmic complexity, standard arguments show that it is undecidable \cite{Kolmogorov:1968thr,ChaitinAC95}.
So the idea we outlined above is really too naive, and we cannot hope for the norm under $V$ to precisely diagnose complexity, as the norm of a vector after a linear transformation is an eminently computable function.

Instead, we must refine our expectations about what information the norm of a state under $V$ can give us, and there is a natural way to do so.
We could seek instead to show that a very small or large norm of a state after application of $V$ implies exponential complexity, but an unchanged norm reveals no information about the complexity of the state.
So there may be exponentially complex states whose norm is quite well-preserved by $V$, but we want to argue that all states with norms that change drastically must have exponential complexity.
This could be done by giving an upper bound on the change in microscopic norm after application of an elementary gate on $\cH_b$.
This is a much more limited sense in which complexity would be relevant, although it would be strong enough to imply the main statement in \cite{Akers:2022qdl} concerning the exponential complexity of null states. 

Unfortunately, there is an immediate obstruction to defining a universal gate set on $\cH_b$ since this space does not have a tensor product structure. 
This means we cannot use the standard qubit model of computation.
Instead, we could select a set of gates which may act on any subspace of $\cH_b$ of the appropriate dimension in the equilibrium basis.
Then attempting to derive a bound on the change in overlaps like \eqref{eq:overlap-bound} for the specific set of subexponential states, as measured in \eqref{eq:relative-complexity}, would be well-defined even for a specific dictionary.
It would be very interesting to understand computation in this model in more detail, and to explicitly verify if subexponential complexity does indeed characterize a subset of $S$-states in a particular member of our non-isometric code ensemble.

\subsection{Fundamental averaging}\label{sec:averaging}

We have studied the error correction properties of the microcanonical holographic dictionary for dilaton gravity.
But what we have really proven is that the typical member of the ensemble of quantum theories dual to these dilaton gravity theories has these properties.
This leads to the question of whether the hallmarks of gravity, such as Lorentz and diffeomorphism invariance, are really present in the typical theory we have studied.

If we are only concerned with generic properties of quantum gravity like Lorentz or diffeomorphism invariance, it is likely that the typical member of the dilaton gravity ensemble is dual to a microscopic bulk theory which has these properties.
An interacting bulk theory with strong interior dynamics can be constructed as a dual to a single instance of the brane states \cite{Blommaert:2021etf}, so the construction in \cite{Blommaert:2021etf} could be taken as the bulk dual of a single member of the dilaton gravity ensemble with branes (eliminating the Hamiltonian integral again by using the microcanonical ensemble).
This theory does have a gravitational description which appears to be diffeomorphism invariant. 
The interpretation of the brane interactions in the original dilaton gravity theory is a bit murky, but the minimal string perspective reviewed in \cite{Blommaert:2021etf} might be helpful in understanding single instances of the dictionary ensemble \eqref{eq:dictionary-3} in more detail.
Of course, since it may be that any particular dictionary can be realized by microcanonical sectors of the specific theories in \cite{Blommaert:2021etf}, our conclusions about interior reconstruction will not hold for arbitrary theories.
Rather, they will hold only for certain theories with high probability as argued in the deviation bound \eqref{eq:overlap-bound} and explained in more detail in Section~\ref{sec:relation}.

We can also try to use indirect evidence to support our assertion.
This indirect evidence comes from situations where the bulk dual of individual ensemble members is known explicitly.
This occurs in e.g.~the $U(1)$ gravity models \cite{Afkhami-Jeddi:2020ezh,Maloney:2020nni} or the $S$-duality ensemble for $\mathcal{N}=4$ super-Yang-Mills theory \cite{Collier:2022emf}, where the individual members of the ensembles are known to be dual to either Chern-Simons theory or type IIB string theory in the bulk, respectively.
Such theories are individually Lorentz and diffeomorphism invariant.
If we approach the issue from a different angle and try to trim down the dilaton gravity ensemble by adding extra bulk degrees of freedom, we are led to the nonlocal brane interaction ideas of \cite{Blommaert:2021gha}.\footnote{The ideas of \cite{Johnson:2022wsr} may also be useful here, but the bulk picture is a bit harder to understand.}
These again are individual quantum theories that are nonlocal but Lorentz and diffeomorphism invariant.
All of these observations give us some confidence that there is really something gravitational about the typical code in \eqref{eq:dictionary-3}.

\subsection{Gravitational operators}

We have considered in this work only abstract bulk operators which shuffle the brane flavor sectors.
This suffices to get an interesting reconstruction theory for our non-isometric code analogous to the one in \cite{Akers:2022qdl}, and roughly corresponds to local unitary operators acting in the interior. 
However, it would be nice to reconstruct a gauge invariant operator with an explicit gravitational interpretation in the Hilbert space of dilaton gravity.
Here we outline how one might approach this problem.

Because dilaton gravity can be canonically quantized on an interval Cauchy slice with one asymptotic and one brane boundary \cite{Gao:2021uro}, we may define nonlocal operators with support on the entire Cauchy slice.
These are, in a sense, well-defined to all orders in a perturbative $G_N$ expansion, and we may try to reconstruct these instead of local interior operators, which are not gauge invariant without gravitational dressing.\footnote{We could also try to use dressed local operators. 
These have been constructed in JT gravity in a diffeomorphism invariant manner \cite{Harlow:2021dfp}, but seem to require dynamical matter fields which may complicate the microscopic theory.}
Such nonlocal operators can in fact be formulated in such a way that they act only within the pure gravity Hilbert space.
This allows a non-perturbative calculation to all orders in $e^{-1/G_N}$ via the Euclidean gravity theory, where all possible Cauchy slices on which the operator could be placed are summed over in the path integral, just as the geometries and topologies themselves are summed over \cite{Saad:2019pqd,Iliesiu:2021ari}.

We define the nonlocal operator
\begin{equation}
    \cO_\Delta \equiv \int \d \ell\; e^{-\Delta \ell} \ket{\ell} \bra{\ell}_\br\ ,
\end{equation}
where $\ket{\ell}_\br$ is the renormalized geodesic length basis of dilaton gravity \cite{Harlow:2018tqv,Gao:2021uro} for the single-sided Hilbert space $\cH_\br$.
We must determine its action on the bulk semiclassical Hilbert space $\cH_b$.
Since we have included only a single state from each flavor copy of $\cH_\br$ in $\cH_b$, the operator $\cO_\Delta$ has a representation on the code space which is diagonal and determined by a single real number
\begin{equation}
    \omega (E,\Delta) \equiv \bra{\textbf{\br}(E)} \cO_\Delta \ket{\textbf{\br}(E)}\ ,
\end{equation}
and the semiclassical unitary operator $\cW_\Delta(g)$ we will try to reconstruct is a phase times the identity on $\cH_b$:
\begin{equation}
    \cW_\Delta \equiv \exp(-\i g\; \omega(E,\Delta))\ I_b\ ,
\end{equation}
where we have introduced the operator coupling $g$.
Evaluating the matrix elements $\bra{E}\cO_\Delta \ket{E'}$ will require an explicit form for the overlaps $\inner{\ell}{E}_\br$, and these were computed in JT gravity in \cite{Gao:2021uro}.

This phase, however, is easily reconstructible on the physical Hilbert space by a unitary operator which is also proportional to the identity and has a matching coefficient $\exp (-\i g \omega)$.
To get something more nontrivial, we should include more states from each flavor copy of $\cH_\br$.
This will extend the code space $\cH_b$ and allow $\cO_\Delta$ to have a nontrivial action, mixing states in $\cH_b$.
Perhaps combining $\cO_\Delta$ with an abstract interior operator $W_b$ mixing the brane flavor sectors would lead to an interesting reconstruction theory.
For a dynamical matter field which extends the pure gravity Hilbert space, avoiding ultraviolet divergences associated with matter loops on higher topologies will be crucial for defining such a theory non-perturbatively, and the results of \cite{Lin:2022rbf,Jafferis:2022wez} could be helpful in this regard.

\subsection{Breakdowns of bulk effective field theory}

One of the primary reasons for the rather rapid development of the entropy formulas and information-theoretic bulk reconstruction we reviewed in Section~\ref{sec:intro} was the observation that bulk effective field theory is in tension with smoothness at the horizon of an evaporating black hole beyond the Page time \cite{Almheiri:2012rt}.
Essentially, due to the structure of entanglement needed to satisfy certain quantum mechanical inequalities, one is forced to consider a high energy ``firewall'' at the horizon instead of a smooth region as expected from the equivalence principle.
The non-isometric reconstruction we have studied here is in some sense the most general framework currently available for describing bulk semiclassical physics in the microscopic theory.
So, it is natural to wonder how breakdowns of semiclassical field theory related to firewalls are manifested in the non-isometric framework.

Non-isometric reconstruction, as we saw, allows for semiclassical physics to be valid up to exponentially small (in $\log |B|$) corrections in any $S$-state and for any $S$-operator.
Following \cite{Akers:2022qdl}, we take this to mean that effective field theory is perfectly valid within this class of states, and in order to see breakdowns we must either exit the set $S$ or consider non-$S$-operators.
The additional input our gravity analysis may have on top of the interpretation in \cite{Akers:2022qdl} is related to the question of whether or not firewall states are ambiguous.

In the two-sided thermofield double state, objects in the interior may be affected by acting with a unitary operator on a single side.
This leads to an issue with describing the experience of an infalling observer, as it seems that we may modify the experience of an observer that falls in from one side by acting with a unitary operator on the other side, despite the fact that the microscopic physics experienced by the observer is controlled only by the reduced density matrix on one side.
This is sometimes called the ``frozen vacuum'' problem \cite{Bousso:2013ifa}, where some additional information is needed to specify which states have stress-energy behind the horizon and which do not.\footnote{Trying to resolve the issue by positing that an infalling observer should be represented by a two-sided operator simply shifts the question.  The challenge in that case is to explain how the holographic dictionary makes an observer two-sided when they begin life as a one-sided operator at the asymptotic boundary.  The ambiguity then reappears, as there are many ways that the dictionary could do this, but presumably only one is actually correct.  See \cite{Jafferis:2020ora,Gao:2021tzr,Jafferis:2022toa} for a proposed unambiguous method involving modular theory.}

In our context, a code with a definite set of $S$-states does not have a frozen vacuum because we trust the predictions of bulk physics for these states up to exponential accuracy, and a generic unitary operation will modify $S$ in a manner which only has a semiclassical description if the operation is an $S$-operator.
In the analysis of \cite{Akers:2022qdl}, although the Haar random unitary ensemble was utilized only for calculational purposes, the ensemble itself was unitarily invariant.
While a typical draw from that ensemble will have an unambiguous set of $S$-states, and therefore will not have a frozen vacuum, the ensemble itself does not have a preferred set of states, and it is not clear how to pick such a set once one has a particular code in hand.

In our gravitational analysis, we arrived at an ensemble in Section~\ref{sec:dictionary} which is certainly not invariant under a unitary transformation of $\cH_b$.
This led us to formulate the equilibrium basis, a set of states which are supposed to have manifestly smooth horizons with zero stress-energy away from the end-of-the-world brane.
In these states, then, there are no firewalls.
Indeed, in any $S$-state that is built around the equilibrium basis, there are no firewalls, and the only interior excitations present are those placed by $S$-operators $W_b$ acting on the equilibrium basis.

This, however, leads us to another sort of breakdown of effective field theory.
Namely, there is a problem with using \eqref{eq:overlap-bound} when we try to include too many brane states in the set $S$.
This breakdown occurs independent of our discussion of complexity theory, and relies only on the form of \eqref{eq:overlap-bound}.
End-of-the-world brane states have smooth horizons just like the thermofield double, and this property does not depend on how many flavors of branes we decide to include in the theory.
So, there should not be an issue with the horizon caused by adding more of them to the set $S$.

Despite this, the breakdown implied by \eqref{eq:overlap-bound} is telling us that if we add too many states with a manifestly smooth horizon to our set $S$ of states on which we wish to reconstruct operators, there may be a pair of them which have large nonzero overlap with reasonable probability.
This large overlap is completely invisible from the semiclassical standpoint, where the rules of the gravitational path integral imply that any overlap should be non-perturbatively small in a typical boundary theory.

So rather than a problem at the horizon, this breakdown of effective field theory could be due to effects which are relevant deep in the interior of the black hole.
This issue may be related to the proper incorporation of interior dynamics, and understanding it more carefully may therefore be important for understanding the black hole singularity.
Alternatively, the restriction on the number of states with smooth horizons in the code space may be a hint that firewall states really are typical in the microcanonical Hilbert space of the black hole.

\acknowledgments

We thank Chris Akers, Lampros Lamprou, Onkar Parrikar, Mark Van Raamsdonk, and Felipe Rosso for helpful discussions.
AK is supported by the Simons Foundation through the It from Qubit Collaboration.


\bibliography{refs}
\bibliographystyle{JHEP}

\end{document}